\documentclass[aps,prb,twocolumn,floatfix,preprintnumbers,superscriptaddress,amsmath,amssymb,longbibliography]{revtex4-2}
\usepackage{xcolor,bm,lineno,multirow,times}
\usepackage{physics}
\usepackage{rotating}
\usepackage{verbatim}
\usepackage{graphicx,tabularx}
\usepackage[multiple]{footmisc}
\usepackage[sort&compress]{natbib}
\usepackage[T1]{fontenc}
\usepackage{soul}
\usepackage{dcolumn}
\usepackage{bm}
\usepackage[version=4]{mhchem}
\usepackage{booktabs}
\usepackage{gensymb}
\usepackage{color}
\usepackage[colorlinks=true,urlcolor=blue,linkcolor=blue,citecolor=blue]{hyperref}
\usepackage{floatrow}

\newfloatcommand{capbtabbox}{table}[][\FBwidth]

\preprint{APS/123-QED}

\begin{document}

\title{{Quantum to classical crossover in generalized spin systems --  the temperature-dependent \\spin dynamics of FeI$_2$}}
\author{D. Dahlbom}
\email{david.dahlbom@gmail.com}
\affiliation{Department of Physics and Astronomy, University of Tennessee, Knoxville, TN 37996, USA}
\author{D. Brooks}
\affiliation{School of Physics, Georgia Institute of Technology, Atlanta, GA 30332, USA}
\author{M. S. Wilson}
\affiliation{Theoretical Division and CNLS, Los Alamos National Laboratory, Los Alamos, NM 87545, USA}
\author{S. Chi}
\affiliation{Neutron Scattering Division, Oak Ridge National Laboratory, Oak Ridge, TN 37831, USA}
\author{A. I. Kolesnikov}
\affiliation{Neutron Scattering Division, Oak Ridge National Laboratory, Oak Ridge, TN 37831, USA}
\author{M. B. Stone}
\affiliation{Neutron Scattering Division, Oak Ridge National Laboratory, Oak Ridge, TN 37831, USA}
\author{H. Cao}
\affiliation{Neutron Scattering Division, Oak Ridge National Laboratory, Oak Ridge, TN 37831, USA}
\author{Y.-W. Li}
\affiliation{Theoretical Division and CNLS, Los Alamos National Laboratory, Los Alamos, NM 87545, USA}
\author{K. Barros}
\affiliation{Theoretical Division and CNLS, Los Alamos National Laboratory, Los Alamos, NM 87545, USA}
\author{M. Mourigal}
\affiliation{School of Physics, Georgia Institute of Technology, Atlanta, GA 30332, USA}
\author{C. D. Batista}
\affiliation{Department of Physics and Astronomy, University of Tennessee, Knoxville, TN 37996, USA}
\author{X. Bai}
\email{xbai@lsu.edu}
\affiliation{School of Physics, Georgia Institute of Technology, Atlanta, GA 30332, USA}
\affiliation{Department of Physics and Astronomy, Louisiana State University, Baton Rouge, LA 70803, USA}
\date{\today}

\begin{abstract}
{Simulating quantum spin systems at finite temperatures is an open challenge in many-body physics. This work studies the temperature-dependent spin dynamics of a pivotal compound, FeI$_2$, to determine if universal quantum effects can be accounted for by a phenomenological renormalization of the dynamical spin structure factor $S({\bm q}, \omega)$ measured by inelastic neutron scattering. Renormalization schemes based on the quantum-to-classical correspondence principle are commonly applied at low temperatures to the harmonic oscillators describing normal modes. However, it is not clear how to extend this renormalization to arbitrarily high temperatures. Here we introduce a temperature-dependent normalization of the classical moments, whose magnitude is determined by imposing the quantum sum rule, e.g. $\int d\omega d{\bm q} S({\bm q}, \omega) = N_S S (S+1)$ for $N_S$ dipolar magnetic moments. We show that this simple renormalization scheme significantly improves the agreement  between the calculated and measured $S({\bm q}, \omega)$ for FeI$_2$ at all temperatures. Due to the coupled dynamics of dipolar and quadrupolar moments in that material, this renormalization procedure is extended to classical theories based on SU(3) coherent states, and by extension, to any SU(N) coherent state representation of local multipolar moments.}
\end{abstract}

\maketitle

\section{Introduction}

The computation of dynamical correlation functions of interacting quantum spin systems at arbitrary temperature is an important and largely open problem in quantum many-body theory. These correlation functions play a crucial role in magnetism. For instance, dynamical susceptibilities associated with two-point correlation functions can reveal the nature of the excitations of spin-liquid phases and their instabilities in the proximity to broken symmetry states. Moreover, they are accessible to various spectroscopic and resonance experiments, providing crucial insight and validation tests for theoretical models. Among these different approaches, inelastic neutron and X-ray scattering provide stringent tests as these experiments can reveal the momentum-, energy- and spin-space dependence of the corresponding dynamical susceptibility~\cite{marshall1_1968,lovesey_1984,enderle_2014,boothroyd_2020}. 

State-of-the-art numerical techniques for simulating quantum many-body problems have severe limitations for computing dynamical correlation functions {at arbitrary temperature}. Exact diagonalization techniques~\cite{Gagliano88} are restricted to small clusters, while density matrix renormalization group (DMRG)~\cite{White92,Schollwock05,Hallberg06} is so far only applicable to one-dimensional systems or narrow ribbons of two-dimensional magnets. Tensor networks can deal with 2D systems and the first attempts of computing dynamical spin structure factors based on single-mode approximations at $T=0$ are very recent~\cite{Chi22}. Quantum Monte Carlo (QMC) techniques are restricted to the small set of models free of the sign problem, and even for that class of problems, the computation of dynamical correlation functions is challenging because the analytical continuation of noisy QMC data from the Matsubara domain to real frequencies can lead to significant uncertainties~\cite{Hirsch83,JARRELL96,Sandvik98,Sandvik16,SHAO2023}. Given these limitations, any {\it efficient} numerical technique that can output a {\it good approximation} to the exact {temperature-dependent} dynamical correlation functions is of general interest to the quantum many-body community. 

\begin{figure}[h!]
        \centering
        \includegraphics[width=1\columnwidth]{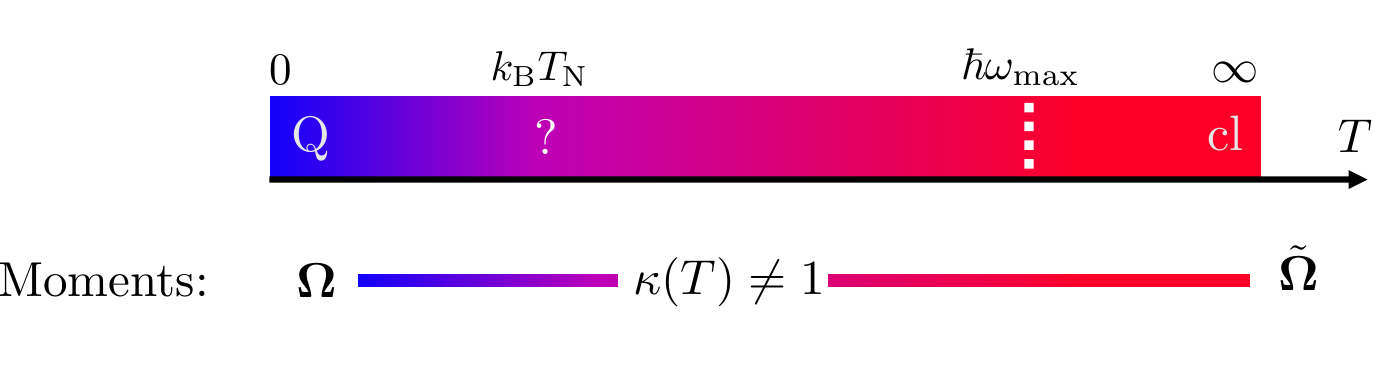}
        \caption{
        Sketch of our approach to simulate the (generalized) dynamical spin structure factor (DSSF) of classical magnetic moments at any temperature, including across a dipolar ordering transition at $T\!=\!T_{\rm N}$, using Landau-Lifshitz dynamics. Our simulations include a renormalization of the classic magnetic moments through the $\kappa(T)$ factor that enforces the dynamical spin structure factor fulfills the quantum-sum zeroth-moment sum-rule at all temperatures, including in the $T\rightarrow\infty$ limit. In this approach the exchange interactions are untouched.}
        \label{fig:1}
\end{figure}

In this work, we show that semi-classical approximations relying on coherent state representations of local quantum moments can fulfill both conditions: their numerical cost is linear in the system size, and the finite-temperature dynamical correlation functions of the quantum many-body system are well approximated by the classical result after applying a well-defined renormalization procedure. To elucidate this procedure, we focus on spin systems where the ground state is approximately a product state between magnetic units. This covers many magnetically ordered systems and can also provide high-fidelity results for spin liquids in a wide range of temperatures. In particular, we detail a general approach to compute the finite-temperature quantum dynamical spin structure factor $\mathcal{S}_{\rm Q}\left({\bf q},\omega\right)$ (see below for precise definitions) typically measured with inelastic neutron scattering. Our approach solves the equation of motion for classical magnetic moments, and while doing so, systematically enforces the quantum sum-rule (``zeroth-moment'') for the dynamical spin structure factor through a temperature-dependent renormalization of the magnetic moments [See Fig.~\ref{fig:1} for a conceptual illustration]. { 
Building on our previous success using the formalism of SU(3) coherent states to model the zero-temperature dynamics of FeI$_2$, a spin-orbital (effectively $S\!=\!1$) magnet~\cite{Bai21,Legros22,Bai23}, we benchmark our method on the same material at finite temperature, obtaining excellent agreement with temperature-dependent neutron scattering results collected across the dipolar long-range ordering transition of the compound at $T=T_{\rm N}$.}

Our work leverages a recent approach~\cite{Zhang21,Dahlbom22,Dahlbom22b} to calculate the classical Landau-Lifshitz (LL) equation of motion of quantum spin systems using coherent states of SU($N$), where $N$ is the number of levels of each magnetic unit ($N\!=\!2S+1$ when the magnetic unit contains only one spin $S$). The resulting generalized classical equations of motion can be linearized at very low temperatures. The corresponding normal modes (e.g., harmonic spin waves for magnetically ordered systems and precessional dynamics for disordered systems) determine the classical dynamical spin structure factor $\mathcal{S}_{\rm cl} \left({\bf q},\omega\right)$. The generalized spin-wave approximation is then obtained by quantizing the harmonic oscillators that describe these normal modes. However, clear qualitative differences exist between the classical and quantum mechanical oscillators. The former have no fluctuations at zero temperature while the latter have zero-point fluctuations. Correspondingly, $ \lim_{T\to 0} \mathcal{S}_{\rm cl} \left({\bf q},\omega\right) =0$ for any $\hbar \omega \neq 0$, while $ \lim_{T\to 0} \mathcal{S}_{Q} \left({\bf q},\omega\right) \neq 0$. In addition, while the detailed balance condition produces a classical dynamical spin structure factor (DSSF) that is symmetric in $\omega$, it gives rise to $\mathcal{S}_{Q} \left(-{\bf q},-\omega\right) = \exp[-\hbar\omega/k_{\rm B} T] \mathcal{S}_{Q} \left({\bf q},\omega\right)$ for the quantum DSSF. Nevertheless, one can use the correspondence principle~\cite{Schofield_1960} between {classical and quantum harmonic oscillators} to obtain $\mathcal{S}_{Q}\left({\bf q},\omega\right)$ as a function of $\mathcal{S}_{\rm cl} \left({\bf q},\omega\right)$ {in the harmonic/linearized approximation}~\cite{Zhang_2019}:
\begin{equation}
    \mathcal{S}_{\rm Q}^{\mathrm{H}}(\mathbf{q}, \omega)=\underbrace{\frac{\hbar\omega}{k_{\rm B} T} \left[1+ n_{\rm B}(\omega/T) \right]}_{g(\omega/T)} \mathcal{S}_{\rm cl}^{\mathrm{H}}(\mathbf{q}, \omega),
\label{eq:ho}
\end{equation}
where $n_{\rm B}(\omega/T) \equiv (e^{\hbar\omega/k_{\rm B} T}-1)^{-1}$ is the Bose function. In essence, the correspondence factor $g(\omega/T)$ accounts for the different probability distributions in the calculation of the two-point correlation function between the classical (Boltzmann, equipartition) and the quantum (Bose-Einstein) harmonic oscillators~\cite{Schofield_1960,Zhang_2019}. For completeness, an intuitive derivation of this factor is provided in Appendix~\ref{sec:q2c-methods}.


Equation~\eqref{eq:ho} underpins the well-known approach of implementing linear spin wave theory (LSWT) by solving the classical linearized LL equations of motion in thermal equilibrium. Taken together, {the LSWT and linearized LL approaches are enormously successful -- perhaps unreasonably so -- in modeling inelastic neutron scattering from magnetically ordered systems and cooperative paramagnets in the low-temperature limit~\cite{Plumb19,Bai19}. Notwithstanding these successes, both approaches break} down at elevated temperatures as they fail to capture the non-linear effects caused by large thermal fluctuations. Since the original LL approach captures these effects at the classical level, it is natural to ask if the classical-to-quantum correspondence principle survives approximately in this non-linear regime. This is the central question our work aims to address and motivates our choice to benchmark it to the spin dynamics of FeI$_2$. Indeed, the dominant on-site term (single-ion anisotropy) in the spin Hamiltonian of FeI$_2$ is expected to have a large effect on thermal non-linearities~\cite{Zhang21,Dahlbom22,Dahlbom22b,Remund22,Pohle23}. This strong influence of the single-ion anisotropy is already visible in the linear regime where  experiments reveal the presence of hybridized dipolar and quadrupolar fluctuations~\cite{Bai21,Legros22,Bai23} requiring the utilization of SU(3) coherent states to faithfully model the dynamical spin structure factor. Recent results on the finite temperature dynamics of the weak-anisotropy compound Ba$_2$FeSi$_2$O$_7$~\cite{Hwan23} suggests that SU(3) coherent states can successfully describe temperature-dependent dynamics provided the norm of classical moments is adjusted between the low- and high-temperature regimes. The results on FeI$_2$ presented here form a strong impetus to extend this work to the strong anisotropy regime where non-linearities in the equation of motion are relevant. 

This paper is organized as follows. Sec.~\ref{sec:dipoles} presents the theory and implementation of a temperature-dependent correction to the dynamics of classical dipoles. Sec.~\ref{sec:multipoles} motivates an extension of the theory to generalized magnetic moments by presenting finite-temperature neutron scattering data on FeI$_2$. We then introduce the formalism 
for implementing the generalized renormalization procedure and provide an explicit benchmark between our approach and the neutron scattering results on FeI$_2$. Sec.~\ref{sec:discussion} discusses our results and concludes the work. Additional computational and experimental details provided in Appendices~\ref{sec:comp-methods} and ~\ref{sec:exp-methods}, respectively. The derivation of the quantum to classical crossover is provided in Appendix~\ref{sec:q2c-methods}.

\section{Renormalized classical dynamics of dipolar moments}
\label{sec:dipoles}

To outline our theoretical approach, we start by considering spin-$S$ moments interacting via a given quantum spin Hamiltonian $\hat{\mathcal{H}}_{\rm Q}$ on a lattice and represented as purely dipolar objects (SU(2) coherent states). We aim at calculating components of the quantum DSSF, which are given in the low-temperature limit by
\begin{equation}
    \mathcal{S}_{{\rm Q}}^{\alpha\beta}(\mathbf{q},\omega) = \sum_{\nu, \mu} \langle \nu | \hat{S}^{\alpha}_\mathbf{q} | \mu \rangle \langle \mu |  \hat{S}^{\beta}_{-\mathbf{q}} | \nu \rangle 
    \frac{e^{-\hbar \varepsilon_{\nu}/k_{\rm B}T}}{\cal Z} \delta(\varepsilon_{\mu} - \varepsilon_{\nu} - \omega),
    \label{eq:sqw}
\end{equation}
where $\epsilon_{\nu}$ and $\vert \nu\rangle$ are eigenvalues and eigenstates of $\hat{\mathcal{H}}_{\rm Q}$ and
\begin{equation}
     \hat{S}^{\beta}_\mathbf{q} = \frac{1}{\sqrt{N_S}} \sum_{j} e^{i {\bm k} \cdot {\bm r}_j} \hat{S}^{\beta}_j.
\end{equation}
The trace over the three diagonal components of the DSSF, $\mathcal{S}_{\rm Q} = \tr_\alpha \left[\mathcal{S}_{\rm Q}^{\alpha\alpha}\right]$, satisfies the ``quantum mechanical'' zeroth-moment sum rule:
\begin{equation}
    \int_{-\infty}^{\infty}d\omega\int d^{d}{\bf q}\, \mathcal{S}_{\rm Q} \left({\bf q},\omega\right)=\sum_{j=1}^{N_S}\left\langle \hat{\mathbf{S}}_{j}^{2}\right\rangle = N_S S\left(S+1\right),\label{eq:qsumrule}
\end{equation}
where $N_S$ is the total number of dipolar moments. In practice, LSWT is the best known approach to (approximately) obtain the low-energy spectrum  of $\hat{\mathcal{H}}_{\rm Q}$ and calculate the quantum DSSF in the harmonic approximation, $\mathcal{S}_{{\rm Q}}^{\rm H}$, using Eq.~\eqref{eq:sqw}.

A corresponding classical theory may be constructed by restricting the possible states
to products of SU($2$) coherent states, $\vert \mathbf{\Omega} \rangle = \bigotimes_j \vert \mathbf{\Omega}_j \rangle$, where $\vert \mathbf{\Omega}_j \rangle$ is an SU(2) coherent state for a spin $S$ at site $j$. 
The classical Hamiltonian corresponding to the above quantum problem is then defined as 
\begin{equation}
\mathcal{H}_{\rm cl} = \lim_{S\rightarrow\infty} \langle \bf{\Omega} \vert \hat{\mathcal{H}}_{\rm Q} \vert\ \bf{\Omega}\rangle,
\label{eq:classical_hamiltonian}
\end{equation}
where the expectation value is taken in the large-$S$ limit.
The SU($2$) coherent state on each site may each be associated uniquely with a classical dipole, $\mathbf{\Omega}_j = \langle \mathbf{\Omega}_j \vert \hat{\mathbf{S}}_j \vert \mathbf{\Omega}_j \rangle$, and the classical Hamiltonian generates the dynamics through the well-known LL equation of motion:
\begin{equation}
\frac{d\mathbf{\Omega}_j}{dt} = \frac{d \mathcal{H}}{d \mathbf{\Omega}_j} \times \mathbf{\Omega}_j.
\label{eq:ll}
\end{equation}
This dynamics generates trajectories from which the classical DSSF may be estimated. Specifically, one samples a spin configuration, $\mathbf{\Omega}_j^0$, at thermal equilibrium as an initial condition, and then calculates a trajectory $\mathbf{\Omega}_j (t)$ where $\mathbf{\Omega}_j(0) = \mathbf{\Omega}^{0}_{j}$. With $\mathbf{\Omega}_{\mathbf{q}}$ denoting the lattice Fourier transform of the trajectory, 
\begin{equation}
{\bm \Omega}_{\bf q} = \frac{1}{\sqrt{N_s}} \sum_{j} e^{i {\bm q} \cdot {\bm r}_j} {\bm \Omega}_{j}
\end{equation}
the classical DSSF is estimated as
\begin{equation}
    \mathcal{S}_{{\rm cl}}^{\alpha\beta}(\mathbf{q},\omega) = \int e^{-i\omega t}
     \left< \Omega_{\bf q}^\alpha(t) \Omega_{-\bf q}^\beta(0)\right> dt 
    \label{eq:classical-sqw},
\end{equation}
where $\alpha, \beta = x, y, z$, 
and the average is taken over many trajectories generated from independent equilibrium samples.
This expression satisfies the ``classical'', zeroth-moment sum rule:
\begin{equation}
    \int_{-\infty}^{\infty} d\omega \int d^d \mathbf{q} \mathcal{S}_{\mathrm{cl}}(\mathbf{q},\omega) = \sum_{j=1}^{N_s} \mathbf{\Omega}_j\cdot\mathbf{\Omega}_j = N_s S^2. 
    \label{eq:csumrule}
\end{equation}

At very low temperatures, the dynamics produced by Eq.~\eqref{eq:ll} is essentially linear and the collective behavior will resemble that of decoupled harmonic oscillators fluctuating about the system's ground state. We can therefore leverage the correspondence factor {$g(\omega/T)$}  in Eq.~\eqref{eq:ho} to obtain a harmonic approximation to the DSSF,
\begin{equation}
    \tilde{\mathcal{S}}_{\rm Q}(\mathbf{q}, \omega)\equiv\frac{\hbar\omega}{k_{\rm B} T} \left[1+ n_{\rm B}(\omega / T) \right] \mathcal{S}_{\rm cl}(\mathbf{q}, \omega).
    \label{eq:quant_equiv_dssf}
\end{equation}
In the low-temperature limit, where the harmonic oscillator
approximation is good, $\tilde{\mathcal{S}}_{\rm Q}(\mathbf{q}, \omega)$ satisfies the quantum mechanical sum rule, Eq.~\eqref{eq:low_temp_sum_rule}, up to corrections of order $S^0$:
\begin{equation}
    \lim_{T\to0} \left[\int_0^\infty\!d\omega \int d^d \mathbf{q}\, \tilde{S}_{\mathrm{Q}}(\mathbf{q},\omega)  \right] = N_s \left(S(S+1) + {\cal O}(S^0) \right).
    \label{eq:low_temp_sum_rule}
\end{equation}
We note, however, that even if we make the strong assumption that the harmonic  approximation holds at higher temperatures, this approach to enforcing the quantum mechanical sum rule will break down as we increase $T$.
In particular, the quantum-classical crossover for each mode of frequency $\omega$ occurs at a temperature $k_B T \approx \hbar\omega$ and the correspondence factor $(\hbar\omega/k_{\rm B} T) \left[1+ n_{\rm B}(\omega / T) \right] \rightarrow 1$ for $k_{\rm B} T \gg \hbar \omega $. Thus, in the high-temperature limit defined as $k_{\rm B} T \gg \hbar\omega_{\rm max}$, where $\omega_{\rm max}$ is the maximum frequency of the normal modes [See Fig.~\ref{fig:1}], the DSSF constructed from classical dynamics using Eq.~\eqref{eq:quant_equiv_dssf} fulfills the classical sum rule, 
\begin{equation}
\begin{alignedat}{2}
    \lim_{T\to\infty} \left[\int_0^\infty\!d\omega \int d^d \mathbf{q}\, \tilde{S}_{\mathrm{Q}}(\mathbf{q},\omega)  \right] = N_s S^2
    \label{eq:high_temp_sum_rule}
\end{alignedat}
\end{equation}
that misses the ${\cal O}(S)$ correction $N_S S$. 

An extra temperature-dependent renormalization is thus required to guarantee that $\tilde{\mathcal{S}}_Q(\mathbf{q}, \omega)$ fulfills the quantum sum rule {\it at any temperature}. As noticed in Ref.~\cite{Huberman08}, the correct quantum sum-rule of $N_s S(S+1)$ can be recovered in the high-temperature limit if the classical dipole moments are renormalized such that $|\mathbf{\Omega}^\prime_j| = \sqrt{S(S+1)}$. In the language of coherent states, this normalization corresponds to the square root of the quadratic Casimir of SU(2),
\begin{equation}
{C}^{(2)}_{\mathrm{SU}(2)} = \hat{S}^\alpha\hat{S}^\alpha = S(S+1).
\label{eq:su2_casimir}
\end{equation}
where from now on we adopt the convention of summation over repeated Greek indices.
This departs from the usual $|{\bm \Omega}_j|=S$ normalization, which guarantees that the classical theory coincides with linear spin-wave theory upon quantization in the zero-temperature limit. As the conflicting sum-rule requirements are independent of the spin Hamiltonian, our approach here is to renormalize the classical dipole moments $\tilde{\bm \Omega}_j (T) = \kappa(T) {\bm \Omega}_j $ where $|{\bm \Omega}_j|=S$, using a temperature-dependent factor $\kappa(T)$.
In the zero-temperature limit, $\kappa(0)=1$, since in this regime the classical-to-quantum correspondence factor {$g(\omega/ T)$} is sufficient to recover the quantum mechanical sum rule up to ${\cal O}(S^0)$ corrections. In the high temperature limit, we set $\kappa(T\to\infty)=\sqrt{S(S+1)}/S=\sqrt{1 + 1/S}$ following Ref.~\cite{Huberman08}. Between these extremes, the value of $\kappa(T)$ may be empirically determined to enforce the quantum mechanical sum-rule of Eq.~\eqref{eq:qsumrule}. 

We emphasize that this $\kappa$ renormalization is solely applied to the normalization of classical moments in the equations of motion, while the Hamiltonian itself is left untouched. Practically, this means that the DSSF of Eq.~\eqref{eq:classical-sqw} is calculated by first sampling equilibrium spin configurations from the classical Hamiltonian, Eq.~\eqref{eq:classical_hamiltonian}, while the dynamics of Eq.~\eqref{eq:ll} are modified by the substitution $\mathbf{\Omega}_j \to \kappa(T)\mathbf{\Omega}_j$. The resulting $\mathcal{S}_{\rm cl}(\mathbf{q},\omega)$ is then multiplied by the classical-to-quantum correspondence factor {$g(\omega/ T)$}  in Eq.~\eqref{eq:ho}.

We observed above that simply applying the correspondence factor to the classical DSSF fails to preserve the quantum sum-rule above the quantum-to-classical crossover temperature. This factor is derived from the correspondence between quantum and classical \textit{harmonic} oscillators. At higher temperatures, however, the harmonic description itself becomes invalid as thermally-induced nonlinearities become relevant. An advantage of the classical approach is that it captures the full nonlinear dynamics of Eq.~\eqref{eq:ll} at no additional computational cost.


\section{Renormalized classical dynamics of generalized moments}
\label{sec:multipoles}

\subsection{Motivation and primer on FeI$_2$}

The above discussion follows the traditional semi-classical treatment of spin Hamiltonians, working with dipolar moments -- coherent states of SU(2) -- in the large-$S$ limit. However, in many quantum materials of current interest, such as our benchmark material FeI$_2$, the low-energy effective Hamiltonian comprises spin $S>1/2$ with (possibly dominant) single-ion anisotropies. In such a case, an appropriate semi-classical treatment needs to consider dipolar and multipolar fluctuations on an equal footing, which calls for the utilization of SU(N) coherent states ($N=2S+1$). As we will show below, our classical moment renormalization procedure extends naturally to this case.

To motivate the need to extend the formalism of Sec.~\ref{sec:dipoles} to larger moments, we first present a brief review on FeI$_2$, the recent understanding of which~\cite{Bai21,Legros22,Bai23} has motivated  theoretical efforts in using generalized coherent states of SU(N) in quantum magnetism~\cite{Zhang21,Dahlbom22,Dahlbom22b}. FeI$_2$ (space group ${\rm P\bar{3}m1}$), belongs to a large family of transition metal dihalides with trigonal symmetry~\cite{McGuire17} comprising perfect triangular-lattice metal layers weakly bonded by van der Waals interactions. The bulk magnetic behavior of FeI$_2$ is characterized by a pronounced easy $c$-axis anisotropy for the effectively $S=1$ moments~\cite{bertrand1974susceptibilite} and the onset of a complex dipolar magnetic order below $T_\text{N}\!=\!9.3$\,K~\cite{gelard1974magnetic}, where $\uparrow\uparrow\downarrow\downarrow$-stripes develop in the triangular plane, breaking the $\bar{3}m$ symmetry. When a magnetic field is applied along the easy-axis, the system undergoes several metamagnetic transitions before reaching saturation around $H_s \approx 12$\,T \cite{fert1973phase}, resulting in a rich temperature-field phase diagram \cite{wiedenmann1988neutron,katsumata2010phase}. 

The excitations of FeI$_2$ have been studied at low temperatures ($T\ll T_{\rm N}$) by various spectroscopic techniques, including far-infrared and time-domain THz~\cite{petitgrand1976far, fert1978excitation, petitgrand1980magnetic, Legros22}, Raman \cite{lockwood1994raman}, electron spin resonance \cite{katsumata2000single, katsumata2000observation}, and neutron scattering \cite{petitgrand1979neutron,Bai21,Bai23}. These experiments show that the dynamical susceptibility of FeI$_2$ is characterized by two types of excitations: conventional single-magnon (SM) modes forming wide bands with dipolar character, and single-ion bound states (SIBS)~\cite{silberglitt1970effect, ono1971two} forming almost-flat bands with quadrupolar character due to the dominant uniaxial anisotropy (see Refs.~\cite{Bai21,Bai23} for cartoons). SIBS are unique to effective $S>1/2$ local Hilbert spaces and are visible in the DSSF of FeI$_2$ because {\it spin-non-conserving} off-diagonal exchange interactions hybridizes them with SM excitations~\cite{Bai21}, [See Appendix \ref{sec:model} for definitions of the model]. As a result, generalized spin-wave theory (using SU(3) coherent states) is a necessary framework to describe the DSSF of FeI$_2$ since it treats both elementary excitations on an equal footing. This hybridization effect also unveils the presence of exchange bound states (non-elementary excitations corresponding to 4- and 6-magnon) \cite{Legros22} as well as strong quantum interactions leading to spontaneous quasi-particle decays induced by magnetic field \citep{Bai23}. These phenomena are non-linear quantum effects that, upon increasing temperature, are expected to crossover into non-linear classical effects. In this work, we focus solely on the spin dynamics in zero magnetic field for which SM and SIBS are the only relevant excitations. {We tune these magnetic excitations with temperature and cross $T=T_{\rm N}$ into the paramagnetic regime. Explaining the temperature-dependent data is a stringent test of our computational approach that requires an extension of the formalism of Sec.~\ref{sec:dipoles} to SU(N) coherent states.}


\subsection{Formalism for multipolar magnetic moments}


The derivation of the traditional classical limit of a spin system begins 
by restricting the quantum state space to products of SU($2$) coherent states
(equivalent to pure two-level states). This is the approach described 
in Section~\ref{sec:dipoles}. When $S=1$, a better classical
approximation can often be derived by starting from a product
of SU($3$) coherent states (equivalent to pure three-level states) \cite{Zhang21}. In
this section, we briefly summarize how to generalize the contents of Section~\ref{sec:dipoles}
to a classical theory based on SU($3$) coherent states.

The DSSF defined in Eq.~\eqref{eq:sqw} involves
correlations among the three spin operators, $\hat{S}^x$, $\hat{S}^y$, and $\hat{S}^z$.
These observables completely characterize the state of an $S=1/2$ spin. In particular,
a one-to-one correspondence may be set up between an SU($2$) coherent state,
$|{\mathbf \Omega}_j\rangle$, and the expectation values of these operators: 
$\langle {\mathbf \Omega}_j \vert \hat{S}^\alpha \vert {\mathbf \Omega}_j \rangle = \Omega^\alpha$.
This is the familiar Bloch sphere construction, and it is this correspondence
that establishes the relationship between the classical dynamics of dipoles in Eq.~\eqref{eq:ll}
and the quantum language of coherent states \cite{Dahlbom22}.

The one-to-one correspondence between dipolar expectation values and SU($2$) coherent
states, $|{\mathbf \Omega}_j\rangle$, does not carry over to SU($3$) coherent states, $|{\mathbf \Psi}_j\rangle$.
This reflects the physical fact that, when $S>1/2$, a spin is characterized not
just by a dipole but also by higher-order multipole moments. 
We may, however, choose a larger set of eight observables, $\hat{T}^\alpha$, including
not just the spin operators but also five quadrupole operators, and establish a
one-to-one correspondence between their expectation values and SU($3$) coherent
states:
$\langle {\mathbf \Psi}_j \vert \hat{T}^\alpha \vert {\mathbf \Psi}_j \rangle = \Psi^\alpha$,
where $1 \leq \alpha \leq 8$.
The arguments of this section essentially repeat those of Section~\ref{sec:dipoles}
while using SU($3$) coherent states and this extended set of observables. In particular,
the generalized DSSF will track correlations among these eight observables instead of
just the spin operators.

Following \cite{Zhang21}, we select the following set of observables, 
\begin{equation}
\begin{alignedat}{4}
& & \quad\quad\hat{T}^{4}_j & =-\left(\hat{S}_j^{x}\hat{S}_j^{z}+\hat{S}_j^{z}\hat{S}_j^{x}\right)\\
\hat{T}^{1}_j & =\hat{S}^{x}_j & 
    \hat{T}^{5}_j & =-\left(\hat{S}^{y}_j \hat{S}^{z}_j+\hat{S}^{z}_j \hat{S}^{y}_j\right)\\
\hat{T}^{2}_j & =\hat{S}^{y}_j & 
    \hat{T}^{6}_j & =\left[\left(\hat{S}_j^{x}\right)^{2}-\left(\hat{S}_j^{y}\right)^{2}\right]\\
\hat{T}^{3}_j & =\hat{S}^{z}_j &
    \hat{T}^{7}_j & =\hat{S}_j^{x}\hat{S}_j^{y}+\hat{S}_j^{y}\hat{S}_j^{x}\\
& & \hat{T}^{8}_j & =\sqrt{3}\left(\hat{S}_j^{z}\right)^{2}-\frac{2}{\sqrt{3}}
\end{alignedat}
\label{eq:su3_generators}
\end{equation}
where $\hat{S}^{x}_j$, $\hat{S}^{y}_j$, and $\hat{S}^{z}_j$ are the three spin operators 
in the standard $S=1$ representation. 
The first three correspond to the components of the dipole
moment and the final five to the components of the quadrupole moment.
We note that these operators constitute a complete set of generators
for the group SU($3$) (a basis for the Lie algebra $\mathfrak{su}(3)$) and satisfy
the orthonormality condition ${\rm tr}[\hat{T}^{\alpha}\hat{T}^{\beta}]=2\delta_{\alpha\beta}$.
With this normalization convention, the generators satisfy a quadratic Casimir,
\begin{equation}
C^{(2)}_{\rm{SU}(3)} = \hat{T}^{\alpha}\hat{T}^{\alpha}=\frac{16}{3}
\label{eq:su3_casimir}
\end{equation}
with  $1 \leq \alpha \leq 8$. This is the natural generalization of the more familiar relation given in Eq.~\eqref{eq:su2_casimir}. 

The DSSF is then expressed in terms of correlations among these observables:
\begin{equation}
    \mathcal{T}_{{\rm Q}}^{\alpha\beta}(\mathbf{q},\omega) = \sum_{\nu, \mu} \langle \nu | \hat{T}^{\alpha}_\mathbf{q} | \mu \rangle \langle \mu |  \hat{T}^{\beta}_\mathbf{q} | \nu \rangle 
    \frac{e^{-\beta \epsilon_{\nu}}}{\cal Z} \delta(\epsilon_{\mu} - \epsilon_{\nu} - \omega),
\label{eq:tqw}
\end{equation}
where $\alpha$ runs over 1 to 8 and
\begin{equation}
     \hat{T}^{\alpha}_\mathbf{q} = \frac{1}{\sqrt{N_S}} \sum_{j} e^{i {\bm k} \cdot {\bm r}_j} \hat{T}^{\alpha}_j.
\end{equation}
As a direct consequence of Eq.~\eqref{eq:su3_casimir}, the trace over the eight diagonal components of the generalized DSSF, $\mathcal{T}_{\rm Q} = \tr_\alpha \left[\mathcal{T}_{\rm Q}^{\alpha\alpha}\right]$, satisfy a sum rule similar to Eq.~\eqref{eq:qsumrule}:
\begin{equation}
    \int_{-\infty}^{\infty}d\omega\int d^{d}{\bf q}\, \mathcal{T}_{\rm Q} \left({\bf q},\omega\right)=\sum_{j=1}^{N_s}\left\langle \hat{\bf T}_{j}^2 \right\rangle = N_s C^{(2)}_{\rm SU(3)}.
    \label{eq:qsumrule_generalized}
\end{equation}

To derive the corresponding classical DSSF, we restrict the quantum state space
to products of SU($3$) coherent states, $|\mathbf{\Psi}\rangle = \bigotimes_j |\mathbf{\Psi}\rangle_j$. The classical Hamiltonian is then defined 
as
\begin{equation}
    \mathcal{H}_{\rm cl} = \langle\mathbf{\Psi} \vert \hat{\mathcal{H}}_{\rm Q} \vert \mathbf{\Psi} \rangle,
    \label{eq:classical_hamiltonian_generalized}
\end{equation}
where we note that, in comparison with Eq.~\eqref{eq:classical_hamiltonian}, no limit appears in Eq.~\eqref{eq:classical_hamiltonian_generalized}. Formally, the limit may be written $\mathcal{H}=\lim_{\lambda_1\to\infty}\langle\hat{\mathcal{H}}\rangle$, where $\lambda_1$ labels degenerate irreps of SU(3). However, after suitable renormalization, $\lim_{\lambda_1\to\infty}\langle\hat{\mathcal{H}}\rangle=\langle\hat{\mathcal{H}}\rangle$. In other words, the classical limit \emph{is} the quantum expectation value, see Ref.~\cite{Zhang21} for details.

As described in \cite{Zhang21, Dahlbom22, Dahlbom22b}, the above classical Hamiltonian generates the dynamics of multipolar moments (three dipolar and five quadrupolar moments) associated with SU(3) coherent states through the generalized LL equation of motion
\begin{equation}
   \frac{d \Psi^\alpha_j}{dt} = f_{\alpha\beta\gamma} \frac{d\mathcal{H}^\alpha}{d\Psi_j^\beta}\Psi_j^\gamma
   \label{eq:ll_generalized}
\end{equation}
where $f_{\alpha\beta\gamma}$ are the structure constants of SU(3), 
\begin{equation}
[\hat{T}^{\alpha}_j, \hat{T}^{\beta}_j] = i f_{\alpha \beta \gamma} \hat{T}^{\gamma}_j,
\end{equation}
and
\begin{equation}
\Psi_j^\alpha = \langle \mathbf{\Psi}_j \vert \hat{T}^\alpha_j \vert \mathbf{\Psi}_j \rangle
\end{equation}
is the $\alpha$-component of the classical multipolar moment at site $j$ and
${\bf \Psi}_j$ a 8-vector of these components.

The generalized classical DSSF may be estimated in a manner directly analogous to 
the approach described in Section~\ref{sec:dipoles}. A sample spin configuration
is generated at thermal equilibrium, ${\mathbf \Psi}_j^0$, and the sample is used as
an initial condition for a trajectory, ${\mathbf \Psi}_j(t)$, generated by the dynamics of Eq.~\eqref{eq:ll_generalized}. Fourier transforming this trajectory on the lattice,
\begin{equation}
{\bm \Psi}_{\bf q} = \frac{1}{\sqrt{N_s}} \sum_{j} e^{i {\bm q} \cdot {\bm r}_j} {\bm \Psi}_{j}
\end{equation}
the generalized classical DSSF may be written,
\begin{equation}
    \mathcal{T}_{{\rm cl}}^{\alpha\beta}(\mathbf{q},\omega) = \int_{-\infty}^{\infty} e^{-i\omega t}
     \Psi_{\bf q}^\alpha(t) \Psi_{-\bf q}^\beta(0) dt
    \label{eq:classical-generalized-sqw},
\end{equation}
and follows the classical sum-rule,
\begin{equation}
    \int_{-\infty}^\infty d\omega \int d^d \mathbf{q} \mathcal{T}_{\mathrm{cl}}(\mathbf{q},\omega) = \sum_{j=1}^{N_s} \mathbf{\Psi}_j\cdot\mathbf{\Psi}_j = \frac{4}{3}N_s
    \label{eq:gencsumrule}
\end{equation}
where  $4/3$ is the normalization  factor of the  8-vector, which may be calculated directly be evaluating $\sum_{\alpha} \langle\hat{T}^\alpha\rangle^2$ for any state. 


{Similar to the purely dipolar case, a temperature-dependent normalization of classical multipolar moments, 
\begin{equation}
   \tilde{\bf \Psi}_j(T) = \kappa(T) {\bf \Psi}_j, \quad |{\bf \Psi}_j|=\sqrt{4/3}\,
\end{equation}
is introduced. It ensures that the quantum sum-rule is saturated for the quantum-equivalent generalized DSSF $\mathcal{\tilde{T}}^{\alpha\beta}_{{\rm Q}}(\mathbf{q},\omega)$ at arbitrary temperature, constructed from the classical DSSF, 
\begin{equation}
    \mathcal{\tilde{T}}_{\rm Q}^{\alpha\beta}(\mathbf{q}, \omega)=\frac{\hbar\omega}{k_{\rm B} T} \left[1+ n_{\rm B}(\omega/T) \right]\mathcal{T}_{\rm cl}^{\alpha\beta}(\mathbf{q}, \omega).
\label{eq:ho2}
\end{equation}}

\subsection{Moment renormalization with SU(3) coherent states}

\begin{figure}[h!]
        \centering
        \includegraphics[width=1\columnwidth]{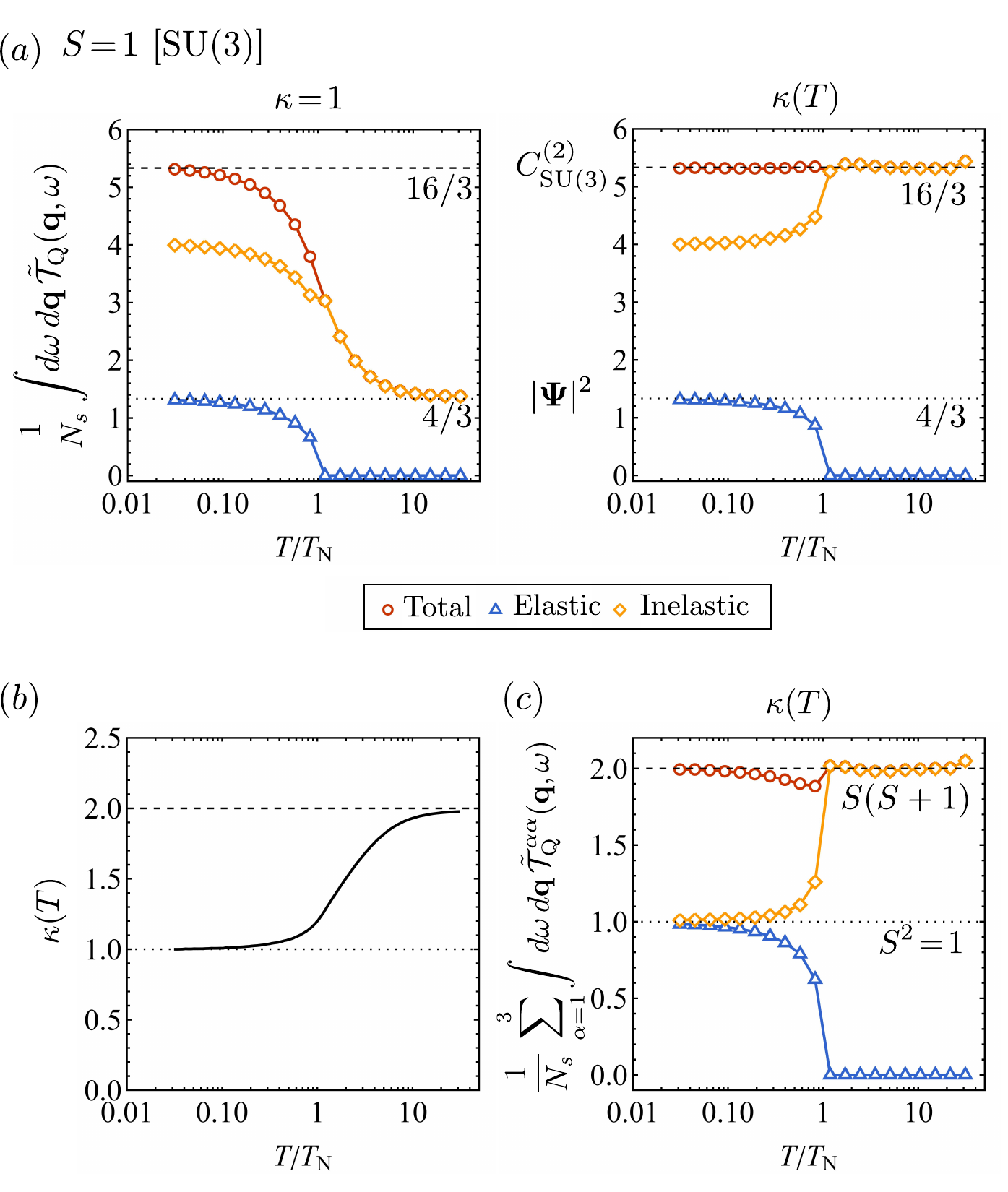}
        \caption{
        (a) Temperature dependence of the generalized DSSF sum-rule for simulations of our FeI$_2$ model (see Appendix~\ref{sec:comp-methods} for details) using SU(3) coherent states ($S\!=\!1$) (left) without, (right) with $\kappa$-renormalization of the 8-component classical magnetic moments (3 dipolar, 5 quadrupolar). Blue triangles, yellow diamonds, and red circles respectively indicate the elastic ($\hbar\omega=0$), inelastic ($\hbar\omega\neq0$) and total contributions to the sum-rule. Throughout, the dotted and dashed lines indicate the classical and quantum sum-rules, respectively. Results are presented in a relative temperature scale $T/T_{\rm N}$, and the sum-rule results are normalized per site. (b) Temperature-dependence of the renormalization factor $\kappa(T)$ obtained from simulations for our FeI$_2$ model. (c) Sum rule restricted to the three dipolar moments of the generalized DSSF. }
        \label{fig:3}
\end{figure}

For concreteness we now illustrate how our approach can be applied in practice with LL simulations for the model spin $S\!=\!1$ Hamiltonian of FeI$_2$ [see Appendix~\ref{sec:comp-methods} for details]. In Fig.~\ref{fig:3}(a) we report the total sum-rule of the generalized DSSF as a function of temperature. As expected, in absence of moment renormalization (i.e. $\kappa\!=\!1$),  the DSSF fullfills the quantum sum-rule in the low-temperature limit with the expected distribution between elastic and inelastic channels for a fully ordered state. Significant deviations are quickly observed in the vicinity to $T \approx T_{\rm N}$ [See Fig.~\ref{fig:3}(a)-left]. Above $T\gtrsim T_{\rm N}$ and in the limit $k_{\rm B}T \approx \hbar\omega_{\rm max}$, the sum rule smoothly approaches the classical result indicating the break-down of the quantum-classical correspondence of Eq.~\ref{eq:ho}. Upon inclusion and optimization of a temperature-dependent renormalization $\kappa(T)$, the quantum sum rule can be conveniently enforced for all temperatures with only minute departures in the vicinity to $T\approx T_{\rm N}$ [See Fig.~\ref{fig:3}(a)-right]. An important observation is that the temperature profile of $\kappa(T)$ [See Fig.~\ref{fig:3}(b)] smoothly interpolates between $\kappa(0)=1$ and $\kappa(\infty) \rightarrow 2$, even in the vicinity of $T_{\rm N}$ . 

The generalized DSSF calculated in Fig.~\ref{fig:3}(a) not only includes dipolar fluctuations accessible to INS but also quadrupolar excitations that are in principle invisible to such probe. As dipolar and quadrupolar fluctuations are treated on equal footing by SU(3) coherent states and can evolve into each other under the LL dynamics, a natural question is if the dipolar components of the generalized DSSF also fulfills a quantum sum rules. Indeed, we observe that the sum-rule restricted to the three dipolar components of the generalized DSSF fulfills expectations for $S=1$ moments represented as SU(2) coherent states, both above and below $T_{\rm N}$ [See Fig.~\ref{fig:3}(c)]. 

\begin{figure}[t]
        \centering
        \includegraphics[width=1\columnwidth]{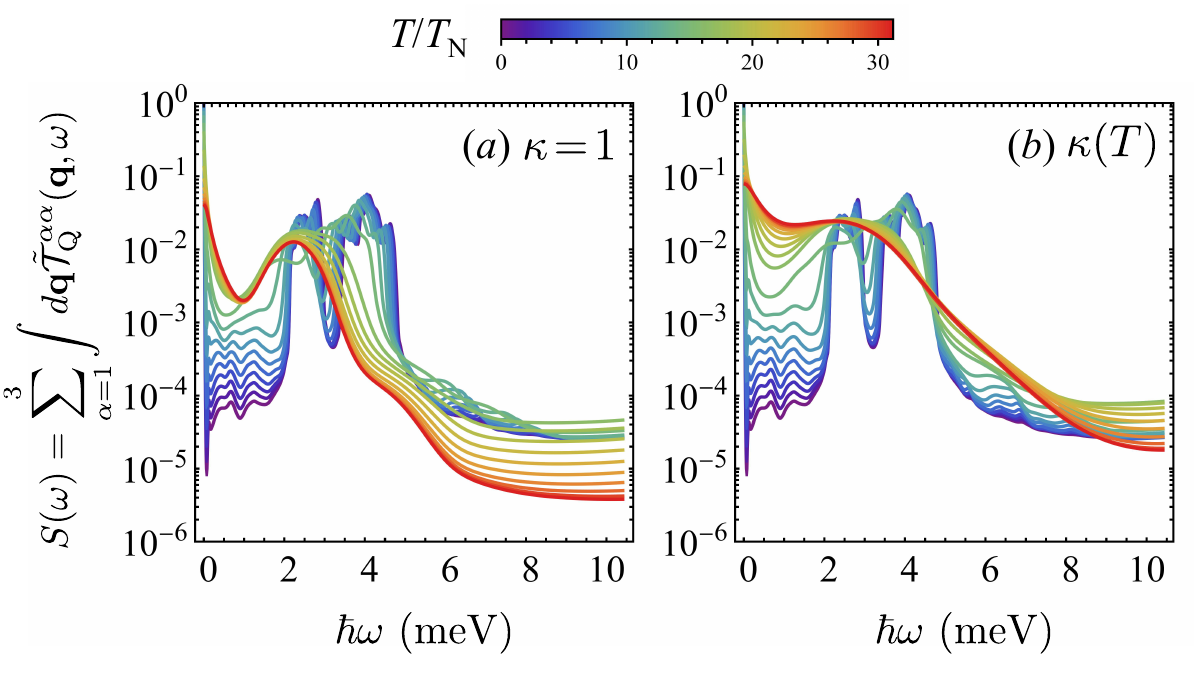}
        \caption{Comparison between the energy-dependence of generalized DSSF between (a) unrenormalized or (b) renormalized moments for the Hamiltonian of FeI$_2$.
        }
        \label{fig:4}
\end{figure}
\subsection{Comparison with experimental temperature-dependent spin dynamics in FeI$_2$}

\begin{figure*}[t]
        \centering
        \includegraphics[width=1\textwidth]{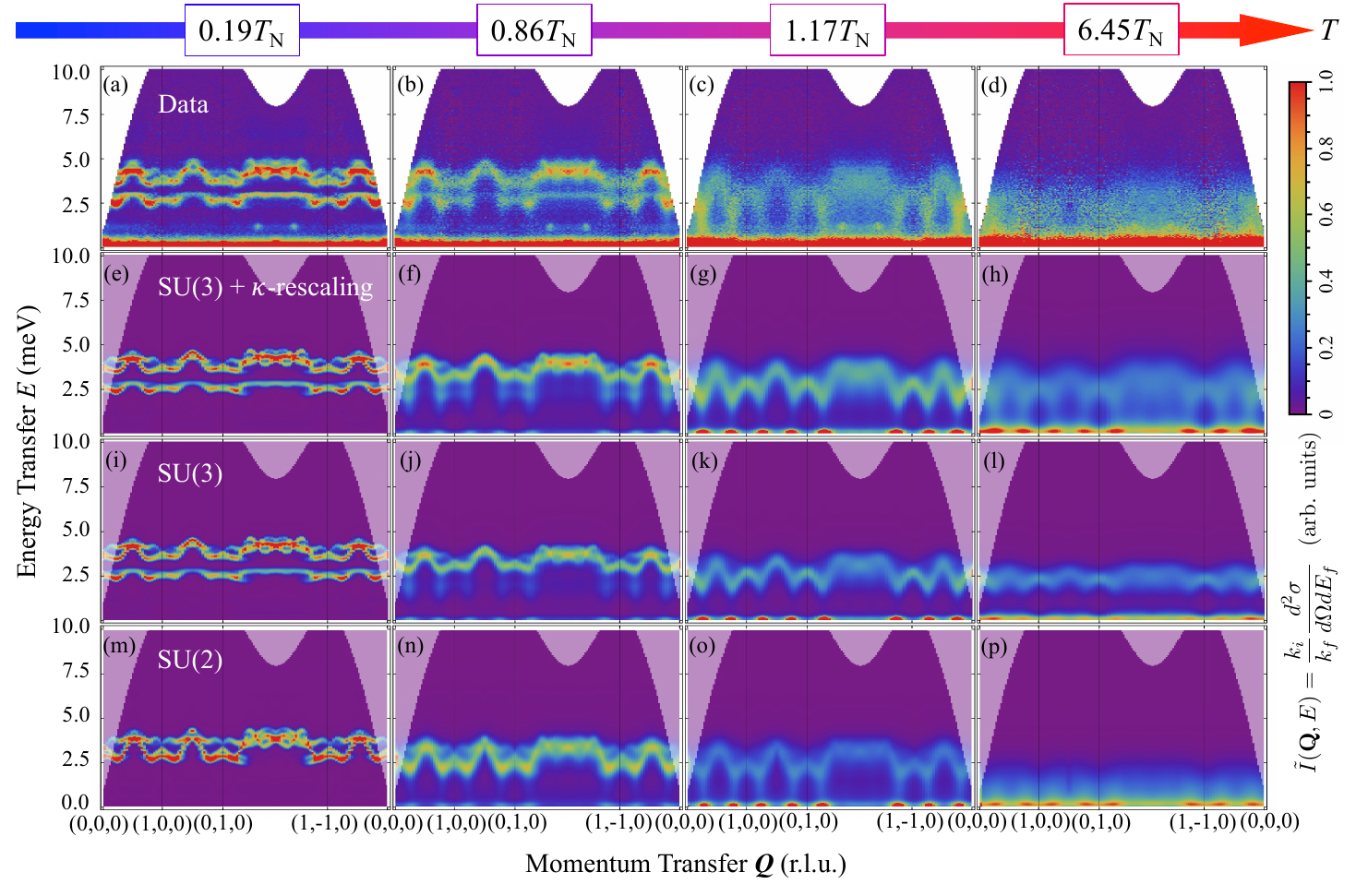}
        \caption{(a-d) Temperature evolution of momentum- and energy-dependent INS intensity $\tilde{I}({\bf Q},E)$ in FeI$_2$ compared to LL simulations of the dipolar DSSF using (e-h) $\kappa(T)$-renormalization for SU(3) coherent states, (i-l) $\kappa = 1$ normalization for SU(3) coherent states, and (m-p) $\kappa = 1$ normalization for SU(2) coherent states. The intensity is shown along chosen high-symmetry directions of the triangular-lattice Brillouin zone, with integration over {$\Delta Q_\ell = 0.05$ \AA$^{-1}$} ($\Delta \ell = \pm 0.1$ r.l.u.) in the out-of-plane direction and {$\Delta Q_\perp = 0.19$ \AA$^{-1}$} in the transverse in-plane direction. The simulated spectra are arranged from the bottom row up with an improved level of agreement with the data. Simulations are performed at relative temperatures $T/T_\text{N}$ matching the relative temperatures of the experimental data. A global intensity scaling factor and an energy-transfer-dependent energy resolution function estimated from the instrument configuration is applied to all simulated results. The two intensity blobs around $E\approx 1$~meV originate from multiple scattering events involving cryostat walls.}
        \label{fig:2}
\end{figure*}

With the quantum sum rule fixed, further benchmarking between simulations and experiments requires inspecting momentum- and energy-resolved neutron scattering results. Before turning to these complete results, we examine the general consequences of the moment renormalization procedure on the energy dependence of dipolar and quadrupolar fluctuations [see Fig.~\ref{fig:4}]. Since the exchange parameters of our Hamiltonian are untouched, our renormalization procedure yields a scaling and re-weighting of the fluctuations in the energy axis but no redistribution of spectral weight as a function of momentum. Above $T=T_{\rm N}$, where the effect of $\kappa(T)$ is the most prominent, the moment renormalization leads to a stretch and enhancement of high-energy fluctuations.

Next, we demonstrate that our approach to solving the generalized LL equations of motion for SU(3) coherent states reproduces all the basic features of the momentum- and energy-resolved experimental data. Furthermore, we show that by incorporating the temperature-dependent renormalization $\kappa(T)$ of classical moments, we achieve a semi-quantitative agreement with the data across the entire temperature range, which crosses the dipolar ordering temperature. For comparison with the experimental INS data, the simulation results are recast [see Appendix~\ref{sec:comp-methods}] as
\begin{equation}
    \tilde{I}_{\rm sim}({\bf Q}, E) = C |f({\bf Q})|^2 \sum_{\alpha,\beta =1}^3 (\delta_{\alpha\beta} - \hat{Q}_\alpha \hat{Q}_\beta) \mathcal{\tilde{T}}^{\alpha\beta}_{\rm Q}({\bf Q},E)
\end{equation}
where $C$ is a constant, $f({\bf Q})$ is the form-factor of Fe$^{2+}$ and the sum is restricted to dipolar components of the generalized DSSF projected perpendicular to ${\bf Q}$.

\begin{figure*}[th!]
        \centering
        \includegraphics[width=0.96\textwidth]{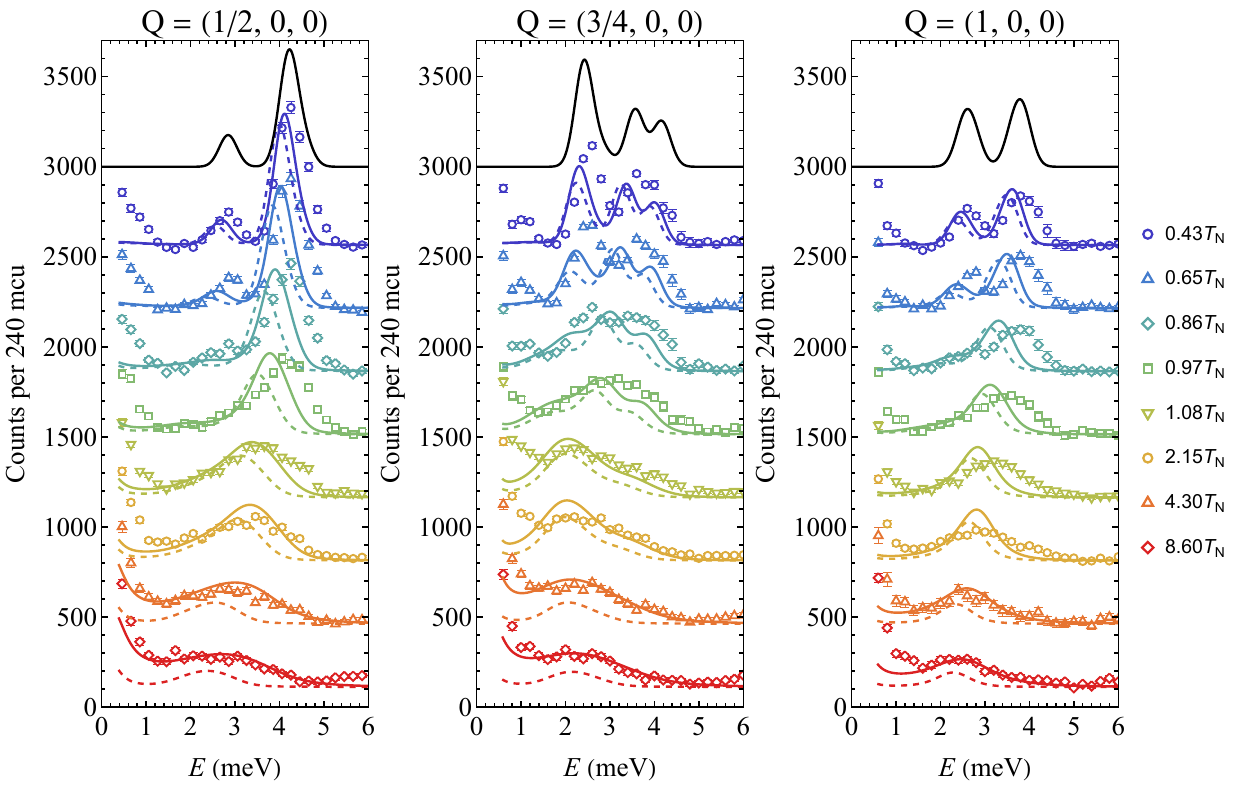}
        \caption{Detailed comparison between INS intensity $\tilde{I}({\bf Q},E)$ in FeI$_2$ (symbols) measured at selected momentum transfers, and corresponding LL simulations of the dipolar DSSF using SU(3) coherent states with (solid lines) and without (dashed lines) $\kappa(T)$-renormalization. A global intensity scaling factor and a Gaussian resolution function (FWHM $=$ 0.47 meV) is applied to all simulated results. Spectra are offset for clarity. The upturn in experimental data above $E \ge 5$\,meV and high temperatures is due to phonons. The spurious signals around 1~meV is due to scattering of higher order neutrons, see Appendix \ref{sec:exp-methods} for further details. The black lines are from generalized linear spin-wave theory calculation. } 
        \label{fig:5}
\end{figure*}

{In Fig.~\ref{fig:2}, we present the temperature evolution of the momentum- and energy-dependent INS intensity $\tilde{I}({\bf Q},E)$ in FeI$_2$ as well as a comparison with our modeling results along several high-symmetry paths in the triangular-lattice Brillouin zone. We have selected experimental temperatures that are representative of four different regimes with respect to the N\'eel temperature $T_{\rm N} = 9.3$\,K: low temperature $T=1.9$\,K same as Ref.~\cite{Bai23} [Fig. \ref{fig:2}(a)], just below and above the dipolar ordering transition $T=8$\,K and $T=11$\,K [Fig. \ref{fig:2}(b--c)], and high temperature $T=60$\,K [Fig. \ref{fig:2}(d)]. 

In the fully ordered phase [Fig.~\ref{fig:2}(a)], the data displays two bands of overlapping magnetic excitations. The upper band, with several branches corresponding to the different sublattices and domains of the dipolar ordered structure, primarily originates from SM excitations, while the lower band originates from SIBS and shows a momentum-dependence due to hybridization with SMs, a phenomenon particularly visible near Brillouin zone centers, such as ${\bf Q}=(1,0,0)$ and ${\bf Q}=(0,1,0)$. Conversely, the flat band along the zone boundary, e.g. between ${\bf Q}=(0,1,0)$ and ${\bf Q}=(1,-1,0)$, has a strong SIBS character. Evidently, these results cannot be captured with traditional dipolar moments [See Fig.~\ref{fig:2}(m)]. Instead, a quantitative match requires the use of generalized magnetic moments represented by SU(3) coherent states [See Fig.~\ref{fig:2}(i)]. In the low-temperature limit, both generalized spin-wave theory ~\cite{Bai21} and generalized classical dynamics [See Fig.~\ref{fig:2}(i)] give identical results, in good agreement with the data.}

The dichotomy of SM and SIBS survives for $T\!=\!0.86~T_{\rm N}$, although excitations broaden significantly [Fig.~\ref{fig:2}(a-b)]. At elevated temperatures [Fig.~\ref{fig:2}(c-d)], as the dipolar order parameter vanishes, the distinction between dipolar and quadrupolar excitations is smeared as $S^z$ is not a good quantum number anymore (even approximately). Just above the phase transition, at $T\!=\!1.17~T_{\rm N}$, the INS data gradually evolves into an energy continuum with momentum modulation reflecting the short-range correlated nature of the underlying dynamic paramagnetic state. In the high-temperature regime, fluctuations become damped (but not quite overdamped), forming an incoherent energy continuum with reduced momentum-dependent correlations. Yet, even in that regime, the bandwidth of the fluctuations matches the energy scale of the single-ion anisotropy, as expected from the results of Ref.~\cite{Zhang21}.

The contrast between the experimental behavior and LL simulations using SU(2) coherent states in the large-$S$ limit is striking. In the ordered phase, despite closely describing the upper excitation band of the data, the lower band is entirely missing [Fig.~\ref{fig:2}(m)]. This is not surprising because although these simulations solve the classical equations of motion using the same {exchange} Hamiltonian parameters, they only consider dipole moments as dynamic degrees of freedom. It is only through simulations with SU(3) coherent states [Fig.~\ref{fig:2}(i)] that both excitation bands can be faithfully captured. 
{Above the phase transition at $T\!=\!1.17~T_{\rm N}$, both SU(2) and SU(3) simulations  [Fig.~\ref{fig:2}(o) and (k)] qualitatively reproduce the momentum-dependent modulation of inelastic signals observed in data. However, the overall intensities are significantly underestimated, with spectral weights predominantly concentrated in the quasi-elastic region. In the high-temperature limit at $T\!=\!6.45~T_{\rm N}$, neither simulation [Fig.~\ref{fig:2}(l) and (p)] provides a good description of the data. The SU(3) simulation [Fig.~\ref{fig:2}(l)] produces a narrow gapped band at the energy of the single-ion anisotropy, which does not agree with the observation. The dipole-only simulations [Fig.~\ref{fig:2}(p)] yield a broad, temperature-damped excitation band that resembles the data but the width of the band is considerably smaller than that of the actual data. }

{By enforcing the quantum sum-rule, the temperature-dependent $\kappa(T)$ renormalization [Fig.~\ref{fig:2}(e-h)] offers two substantial improvements. First, inelastic signals are generally enhanced such that the change of overall intensities as a function of temperature is much more gradual and consistent with the data. In contrast, inelastic intensities drop too quickly above the phase transition in SU(2) simulations and SU(3) simulations without $\kappa(T)$ renormalization. Second, the bandwidth of excitations is increased, which is a result of amplified local mean-fields due to the rescaling of moments. This effect produces a better agreement in energy profiles of INS spectra at all temperatures, particularly important in the high-temperature limit [Fig.~\ref{fig:2}(h)]. A more detailed inspection of the temperature-dependence of the INS intensity at three distinct momenta ${\bf Q}=(h,0,0)$ with $h=1/2, 3/4$ and $1$ is presented in Fig.~\ref{fig:5}. The bandwidth and energy position of the simulated excitation continuum in the high-temperature matches quantitatively with the data.} Overall, this establishes that generalized LL dynamics with temperature-renormalized SU(N) coherent states are a necessary minimal approach to perform semi-classical simulations of the DSSF of quantum spin systems at finite temperatures.

\section{Discussion and Conclusion}
\label{sec:discussion}

{In this section, we discuss some limitations of our approach and the scope for future work.  The comparison at $T\!=\!0.86~T_{\rm N}$ [Fig.~\ref{fig:2}(b) and (f)] reveals an obvious discrepancy because the SIBS excitations are more visible in the data than in the  simulations.  A close inspection of spectra in Fig.~\ref{fig:5} reveals that the simulated SIBS excitations have a weaker intensity than experiments due to a broadened lineshape. More generally, the detailed temperature-dependent comparison also shows that simulations systematically underestimate the energy of the dominant excitations. The discrepancy is around $0.1$~meV at $T\!=\!0.49T_{\rm N}$ and it increases upon approaching $T\!=\!T_{\rm N}$. This is not due to the inaccuracy of our Hamiltonian parameters as generalized linear spin-wave calculations at zero-temperature are in good agreement with the peak positions in the data [black lines in Fig.~\ref{fig:4}].

To shed some light on the source of these discrepancies, we examine the temperature dependence of the order parameter shown in Fig.~\ref{fig:6}. In the experiment, the system undergoes a first-order transition at $T\!=\!T_{\rm N}$ with the Bragg peak intensities saturating within a small temperature window. The simulated order parameter, on the other hand, grows at a much slower rate and only reaches $70\%$ of its maximal value at $T\!=\!0.5~T_{\rm N}$. The corresponding reduction in the average molecular fields due to the excessive amount of thermal disorder in the simulations reduces the bandwidth of the collective modes. In other words, we employ thermalized classical configurations of SU(3) coherent states that fail to describe the magnetic phase transition of FeI$_2$ because they cannot reproduce its first-order nature. This problem is rooted in the Boltzmann statistics of the classical model: the equipartition theorem guarantees a constant specific heat at low temperatures, which vastly overestimates the exponentially suppressed specific heat of our gapped quantum mechanical system. A possible improvement is to incorporate the quantum (Bose-Einstein) statistics through random Langevin-like forces with a specific power spectral density (colored noise)~\cite{Savin12}. This improvement could lead to a first-order thermodynamic phase transition in agreement with the experimental observation. Indeed, our classical simulations of the spin Hamiltonian of FeI$_2$ in the {\it Ising limit} (where Ising variables replace moments in the limit of infinite single-ion anisotropy) produce a first-order thermodynamic phase transition at $T_{\rm N, Ising} = 8.7$~K that is in very good quantitative agreement with the experimental data [Fig.~\ref{fig:6}]. 

\begin{figure}[h!]
        \centering
        \includegraphics[width=1\columnwidth]{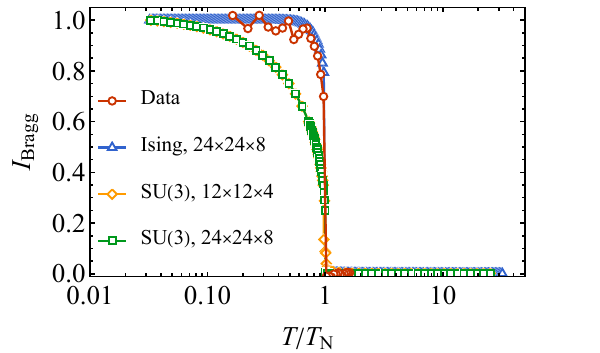}
        \caption{Temperature dependence of the dipolar order parameters in FeI$_2$. {The red circles indicate experimental integrated intensities of a magnetic Bragg peak. The blue triangles are simulated Bragg intensities of an Ising model [See Appendix \ref{subsec:IsingMC} for details], which best captures the first-order nature of the phase transition. The yellow diamonds and green squares are simulations with SU(3) spins and different system sizes. These results show the discrepancy with experimental data is not a finite size effect and is instead rooted in the simulation approach.} }
        \label{fig:6}
\end{figure}

In summary, our work shows that generalized classical LL simulations can quantitatively describe the finite-temperature dynamics of coupled local dipoles and quadrupoles in FeI$_2$. The high numerical efficiency of this approach allows for simulation of the entire temperature-, momentum-, energy-, and spin-dependence of the dynamical spin structure factors with high resolution. The generalized LL dynamics is derived using coherent states of the degenerate representations of SU(3)~\cite{Zhang21,Dahlbom22,Dahlbom22b,Dahlbom23}. 
This approach benchmarks well with inelastic neutron scattering results on FeI$_2$ despite two significant complications. First, this compound hosts hybridized dipolar and quadrupolar fluctuations that must be treated equally. Second, the system orders magnetically, and the experimental spin dynamics need to be simulated over two decades of temperature from $T\!\approx0.1~T_{\rm N}$ to $T\!\approx10~T_{\rm N}$. Even under these constraints, and provided the classical moments are renormalized as a function of temperature so that the DSSF satisfies the quantum sum rule, the entire dynamical response of the material is approximated with good fidelity. The gapped excitation spectrum of FeI$_2$ (in zero magnetic field), and the approximate product state nature of the ground state, aided by the large single-ion anisotropy, are likely favorable conditions. In the future, extensions to gapless systems are planned.

The effectiveness and efficiency of classical and semi-classical  approaches to describe spin dynamics in quantum magnetism is remarkable. Even in the extreme quantum limit of  quantum spin-liquids, whose ground states  cannot be approximated by product states, preliminary results~\cite{Samarakoon17, FrankeKnolle_2022} indicate that the semi-classical approach offers significant benefits and a good approximation of the dynamical response function as long as the temperature is sufficiently high. In general, a unique advantage of classical and semi-classical approaches is that the computational cost of their numerical implementation scales linearly in the number of spins. This property is crucial for solving the inverse scattering problem of extracting models from inelastic scattering data because of the requirement of solving the direct problem for a very large number of candidate spin Hamiltonians. 
}

\begin{acknowledgements}
The work of X.B. at LSU was supported by the Louisiana Board of Regents Support Fund. The work of K.B. was supported by the LANL LDRD program. The work of D.D. and C.D.B. at UTK, and D.B., M.M., and X.B. (earlier work) at GT was supported by the U.S. Department of Energy, Office of Science, Basic Energy Sciences, Materials Sciences, and Engineering Division under award DE-SC-0018660. We thank Tyrel McQueen and Adam Phelan for their help with crystal growth at the National Science Foundation's PARADIM (Platform for the Accelerated Realization,  Analysis, and  Discovery  of  Interface  Materials), funded under Cooperative Agreement NSF-DMR-2039380. This research used resources at the High Flux Isotope Reactor and Spallation Neutron Source, a DOE Office of Science User Facility operated by the Oak Ridge National Laboratory.
\end{acknowledgements}

\appendix

\section{Computational Methods}
\label{sec:comp-methods}
All calculations, excluding the final energy
convolution, were performed using the Sunny.jl package \cite{Sunny}. They may be reproduced using example code available at \cite{CodeExamples}.
\subsection{Calculating Structure Factors}
\label{subsec:calculating_sfs}

The structure factors were calculated by first sampling spin configurations
at thermal equilbrium and, for each of these samples, calculating
a dissipationless trajectory, $\Omega_{j,n}^{\alpha}\left(t\right)$,
where $j$ is the site index, $n$ the sample number, and $\alpha$
spin component. This trajectory was then Fourier transformed,
both on the lattice and in time,
\[
\Omega_{{\bf q}}^{\alpha}\left(\omega\right)=\frac{1}{\sqrt{2\pi N_{s}}}\int_{-\infty}^{\infty}d\omega e^{i 2 \pi \omega t}\sum_{j}e^{i{\bf q}\cdot{\bf r}_{j}}\Omega_{j}^{\alpha}\left(t\right)
\]
where ${\bf r}_{j}$ is the position of site $j$. The convolution
theorem then allows us to estimate the classical DSSF as
\begin{equation}
\mathcal{S}^{\alpha\beta}_{\rm cl}\left({\bf q},\omega\right)=\frac{1}{N_{{\rm samples}}}\sum_{n}\Omega_{{\bf q},n}^{\alpha}\left(\omega\right)\Omega_{{\bf -q},n}^{\beta}\left(-\omega\right).
\label{eq:dssf_estimate}
\end{equation}
The final estimate of the quantum DSSF was calculated using the classical-to-quantum correspondence factor of Eq.~\eqref{eq:ho}. 

All simulations were performed numerically on a lattice of dimensions
$24\times24\times8$ and in discrete time, so in fact the discrete
analog of the above equation was the actual computation. The Fourier
transform was performed using the FFTW package.

For the traditional Landau-Lifshitz simulations, the results of which
are shown in the final row of Fig.~\ref{fig:2}, $\Omega_{j}^{\beta}\left(t\right)$
was calculated by numerically simulating Eq.~\eqref{eq:ll} directly. For the
SU$(3)$ results, the generalized dynamics of Eq.~\eqref{eq:ll_generalized} were simulated
in an equivalent Schrödinger formulation \cite{Dahlbom22}. This approach
evolves an SU(3) coherent state, $\vert{\bf \Psi}_{j}\text{\ensuremath{\left(t\right)}\ensuremath{\ensuremath{\rangle}}}$,
on each site, from which the purely dipolar trajectory may be recovered
by taking expectation values: $\Psi_{j}^{\alpha}=\langle{\bf \Psi}_{j}\left(t\right)\vert\hat{S}^{\alpha}\vert{\bf \Psi}_{j}\left(t\right)\rangle$.
These trajectories were Fourier transformed as above, yielding $\Psi_{{\bf q}}^{\alpha}\left(\omega\right)$.
The DSSF was calculated as in Eq.~\eqref{eq:dssf_estimate}, with the substitution $\Omega_{{\bf q}}^{\alpha}\left(\omega\right)\to\Psi_{{\bf q}}^{\alpha}\left(\omega\right)$.

Equilibrium samples were generated by first initializing the system
in one of the three degenerate ground states (ordering wave vectors ${\bf Q}=\left(0,-1/4,1/4\right)$,
$\left(1/4,0,1/4\right)$, and $\left(-1/4,1/4,1/4\right)$). The
system was then thermalized using Langevin integration \cite{Dahlbom22}
at the desired temperature until reaching equilibrium, as determined
by examining the ergodicity of energy trajectories. A time step of
$\Delta t=0.004$ meV$^{-1}$ was used to ensure numerical stability,
and it was found that a thermalization duration of $12.5$ meV$^{-1}$
was sufficient to reach equilibrium at all temperatures examined.
The coupling to the thermal bath was determined by a phenomenological
parameter, $\lambda$, which was set to $0.1$ -- see \cite{Dahlbom22} further
details.
For the results presented in Figs. \ref{fig:3} and \ref{fig:4}, a total of 1200 samples were
collected at each temperature. As a result of the ground state degeneracy,
the experimental data effectively averages over each of the possible
ground states due to domain formation. Therefore, of the 1200 samples,
400 were thermalized starting from each of the degenerate ground states.

The dissipationless trajectories were calculated using a symplectic
integration scheme to ensure energy conservation \cite{Dahlbom22}. A time
step of $\Delta t=0.025$ was selected to ensure numerical stability,
and trajectories were run for a duration of $180$ meV$^{-1}$. Only
every 12th step of the trajectory was recorded, resulting in a maximum
resolved energy of $\hbar\omega=\pi/(12\Delta t)=10.47$ meV, with
300 non-negative energy bins of width $\hbar\Delta\omega=0.03496$ meV.
Finally, the $\omega$-axis of the resulting data was convolved with
an $\omega$-dependent smoothing kernel based on the energy resolution
of the SEQUOIA instrument.

\subsection{Estimating the spin renormalization factor $\kappa\left(T\right)$}

Estimating the $\kappa\left(T\right)$ requires computation of the
generalized DSSF for SU(3) coherent states Eq.~\eqref{eq:classical-generalized-sqw}. The procedure
for calculating this is identical what was described for the ordinary
DSSF in the SU(3) case, involving the sampling of initial condition
from thermal equilibrium and simulating dissipationless trajectories.
The main addition is that the dissipationless trajectories now include
the expectations not only of the dipole operators but also the remaining
generators given in Eq.~\eqref{eq:su3_generators}, $\Psi_{j,n}^{\alpha}=\langle{\bf \Psi}_{j}\left(t\right)\vert\hat{T}^{\alpha}\vert{\bf \Psi}_{j}\left(t\right)\rangle$,
with $\alpha$ ranging from 1 to 8. Using these trajectories, the
generalized, classical DSSF is estimated as,
\[
\mathcal{T_{\rm cl}}^{\alpha\beta}\left({\bf q},\omega\right)=\frac{1}{N_{{\rm samples}}}\sum_{n}\Psi_{{\bf q},n}^{\alpha}\left(\omega\right)\Psi_{{\bf -q},n}^{\beta}\left(-\omega\right).
\]
When performing the $\kappa$ rescaling, this same procedure is used,
except that the dynamical trajectories, $\Psi_{j}^{\alpha}\left(t\right)$,
are calculated using renormalized spins. In other words, the dynamics
of Eq.~\eqref{eq:ll_generalized} are subjected to the substitution $\Psi_{j}^{\alpha}\to\kappa\Psi_{j}^{\alpha}$. How well the DSSF calculated using the renormalized dynamics, $\mathcal{T}_{\kappa} = \rm{tr}[\mathcal{T}_{\kappa}^{\alpha\beta}]$, satisfies the quantum sum rule
may be determined by evaluating
\begin{equation}
\int_{-\infty}^{\infty}d\omega\int d^{d}{\bf q}\hspace{3mm} g\left( \omega/T \right)\mathcal{T_{\kappa}}\left({\bf q},\omega\right),
\label{eq:kappa_sum}
\end{equation}
where $g\left(\omega/T\right)$ is the classical-to-quantum correspondence factor of Eq.~\eqref{eq:ho}.

Using a binary search algorithm, $\kappa$ was estimated for 100 logarithmically-spaced temperatures, $T_i$, between 0.1 and 100.0 K. For each temperature, the bounds of the $\kappa$ search space were set to $\kappa_{\rm lo}=\kappa(0)=1.0$ and $\kappa_{\rm hi} = \kappa(\infty) = 2.0$. An initial
guess was chosen between these two values, $\kappa_i^0$, where the upper index is the iteration number and the lower index corresponds 
to the given temperature, $T_i$. The DSSF was then calculated using $\kappa_i^0$, and the resulting sum, Eq.~\eqref{eq:kappa_sum}, was
evaluated. If the sum exceeded the reference quantum sum, $N_{s}C_{{\rm SU}\left(3\right)}^{(2)}=\frac{16}{3}N_{s}$, with $N_s$ the number of sites, then the process was repeated, setting the new candidate to $\kappa_i^1 = (\kappa_{\rm lo} + \kappa_i^0)/2$ and resetting the upper bound to $\kappa_{\rm hi} = \kappa_i^0$. If the sum was less than $\frac{16}{3}N_{s}$, then the new candidate was taken to be $\kappa_i^1 = (\kappa_i^0 + \kappa_{\rm hi})/2$ the lower bound was reset as $\kappa_{\rm lo} = \kappa_i^0$. This process was repeated until the estimated sum satisfied the condition
\[
\left|\int_{-\infty}^{\infty}d\omega\int d^{d}{\bf q} \hspace{3mm} g\left(\omega/T\right)\mathcal{T}_{\kappa^{n}_{i}}\left({\bf q},\omega\right)-N_{s}C_{{\rm SU}\left(3\right)}^{\left(2\right)}\right|<0.01.
\]
$\kappa_i$ was then defined as $\kappa_i^n$. $\kappa_0^0$ was set to $1$ initially, since the renormalization for the lowest temperature was expected to be very close to $\kappa(0)=1$. For each successive temperature, the initial guess was set to the value determined for the next lowest temperature, i.e., $\kappa_i^0=\kappa_{i-1}$, as $\kappa(T)$ was expected to vary smoothly with $T$. The results of this procedure are
presented in Fig.~\ref{fig:3}.

\subsection{Simulated N\'eel temperature and temperature rescaling}
In general, classical simulations will not reproduce the N\'eel temperature, $T_N$, of the corresponding quantum problem. It is therefore necessary
to rescale the simulation temperature, $T_{\rm sim}$, to make a meaningful comparison with a corresponding experimental temperatures, $T$. In the work presented here, the experimental temperature was rescaled linearly by the ratio of the simulated N\'eel temperature, $T_{\rm N, sim}$, to the experimental one, $T_{\rm N}$:  $T_{\rm sim} = \left(T_{\rm N, sim}/T_N\right)T$.

To determine $T_{\rm N, sim}$, the dipolar order parameter was estimated at 88 
logarithmically-distributed temperatures between $0.1$ and $77.0$ Kelvin, with a greater
density of sampling in the transition region, as illustrated in Fig.~\ref{fig:6}. For each temperature, 3000 equilibrium samples were drawn, $\Omega^\alpha_{j,n}$, where $\alpha$ is the spin component, $j$ is the site index, and $n$ the sample number. Samples were generated using the Langevin integration scheme described in Sec.~\ref{subsec:calculating_sfs}. Specifically, the system was initialized in the ground state with ordering wave vector $\mathbf{q}_{\rm ord} = (0,-1/4,1/4)$. The Langevin dynamics were run to until the system reached equilibrium, as determined by the ergodicity of the time series generated by the system's energy. A duration of 15\,meV$^{-1}$ was sufficient for all temperatures. Subsequent samples were generated by running the dynamics for intervals long enough to decorrelate the time series generated by the systems' energy. Decorrelation times ranged from 12\,meV$^{-1}$ at low temperature to 0.8\,meV$^{-1}$ at high temperatures.

Each sample was Fourier transformed on the lattice,
\begin{equation}
\Omega_{{\bf q}, n}^\alpha = \frac{1}{\sqrt{N_s}}\sum_j e^{i\mathbf{q}\cdot\mathbf{r}_j}\Omega^\alpha_{j,n}
\end{equation}
and the instantaneous structure factor for a given temperature $T$ was calculated as 
\begin{equation}
\mathcal{S}_T^{\alpha\beta}\left({\bf q}\right)=\frac{1}{N_{{\rm samples}}}\sum_{n}\Omega^{\alpha}_{{\bf q},n}\Omega^\beta_{{\bf -q},n}.
\label{eq:instant_sf}
\end{equation}
The Bragg intensity, 
\begin{equation}
    I_{\rm Bragg}\left(T\right) \equiv \sum_{\alpha} \mathcal{S}_T^{\alpha\alpha}\left(\mathbf{q}_{\rm ord}\right)
\end{equation}
was determined by calculating the trace of the instant structure factor at the ordering wave vector. $T_{\rm N, cl}$ was defined as the temperature at which a discontinuity appeared in the numerical derivative of the $I_{\rm Bragg}(T)$. For the SU(2) (traditional Landau-Lifshitz) simulations, $T_{\rm N, cl}=5.35$\,K; for the SU(3) simulations, $T_{\rm N, cl}=3.05$\,K.

\subsection{Monte Carlo simulations of the Ising model}
\label{subsec:IsingMC}

With the same $J^{zz}$ exchange interactions from Tab.~\ref{tab:exchange}, we simulated a classical Ising model where $D$ was replaced by the constraint that spin dipoles be allowed only to point in the $\pm z$ directions. 
We utilized the parallel tempering \cite{PT_swedensen} Monte Carlo method with single spin flips to thermalize the system and generate equilibrium samples.
Using a parallel tempering simulation with 256 logarithmically-distributed temperature points between 0.1 and 77.0 Kelvin, the average dipolar order parameter was determined from Eq. \ref{eq:instant_sf} using 100 independent equilibrium samples at each data point. 
Results for the described simulation are shown in Fig. \ref{fig:6}.
%

\subsection{Hamiltonian model of FeI$_2$}\label{sec:model}
The Hamiltonian model of FeI$_2$, Eq.~\eqref{eq:FeI2_ham}, consists of three intra-layer, three inter-layer interactions and a uniaxial single-ion anisotropy, [See Fig.~\ref{fig:exchange} and Tab.~\ref{tab:exchange}].
For nearest-neighbor bonds, all symmetry-allowed diagonal and off-diagonal exchange interactions are included in the model. For further-neighbor bonds, only diagonal anisotropy is considered. We adopt the representative values of model parameters, Tab.~\ref{tab:exchange}, obtained in Ref.~\cite{Bai21} by joint fits to the energy-integrated paramagnetic diffuse scattering data and the energy-resolved magnetic excitation data in the ordered phase.

\begin{widetext}
\begin{equation}\label{eq:FeI2_ham}
\begin{split}
     \quad\text{nearest neighbor} &\left\{\begin{aligned}
        &\left.\mathcal{H} = \sum_{\left\langle i,j\right\rangle }\big[J_{1}^{zz}{S}_{i}^{z}{S}_{j}^{z}+{\frac{1}{2}}J_{1}^{\pm}\left({S}_{i}^{+}{S}_{j}^{-}+{S}_{i}^{-}{S}_{j}^{+}\right)\right\}\quad \text{spin-conserving}\\
        & \left.\begin{split}
        & \quad\quad\quad+ {\frac{1}{2}}J_{1}^{\pm\pm}\left(\gamma_{ij}{S}_{i}^{+}{S}_{j}^{+}+\gamma_{ij}^{*}{S}_{i}^{-}{S}_{j}^{-}\right) \\
        & \quad\quad\quad-\dfrac{iJ^{z\pm}_{1}}{2}\left[ (\gamma^{*}_{ij}{S}^{+}_{i}-\gamma_{ij}{S}^{-}_{i}){S}^{z}_{j}+ {S}^{z}_{i}(\gamma^{*}_{ij}{S}^{+}_{j}-\gamma_{ij}{S}^{-}_{j})\right]\big] 
        \end{split}\right\}\quad \text{spin-non-conserving}
      \end{aligned}\right. \\
    & \quad\text{ } \underbrace{ \quad +\sum_{\left( i,j\right) }\big[J_{{\rm f.n.}}^{zz}{S}_{i}^{z}{S}_{j}^{z}+{\frac{1}{2}}J_{{\rm f.n.}}^{\pm}\left({S}_{i}^{+}{S}_{j}^{-}+{S}_{i}^{-}{S}_{j}^{+}\right)\big]}_{\text{further neighbor}} \underbrace{-D\displaystyle{\sum_i} (S_i^z)^2}_{\text{single-ion}}\, 
\end{split}
\end{equation}
where $\gamma_{ij}=e^{i\theta_{ij}}$ are bond-dependent phase factors with $\theta_{ij}=\theta_{ji}=0,+\frac{2}{3},-\frac{2}{3}$ depending on the direction of the bond of the triangular lattice~\cite{Maksimov_2019}.

\thisfloatsetup{style=plain,capposition=bottom, heightadjust=object}
\begin{figure}[th!]
\CenterFloatBoxes
{\begin{floatrow}
{\ffigbox[6cm]{%
  \includegraphics[width=0.9\columnwidth]{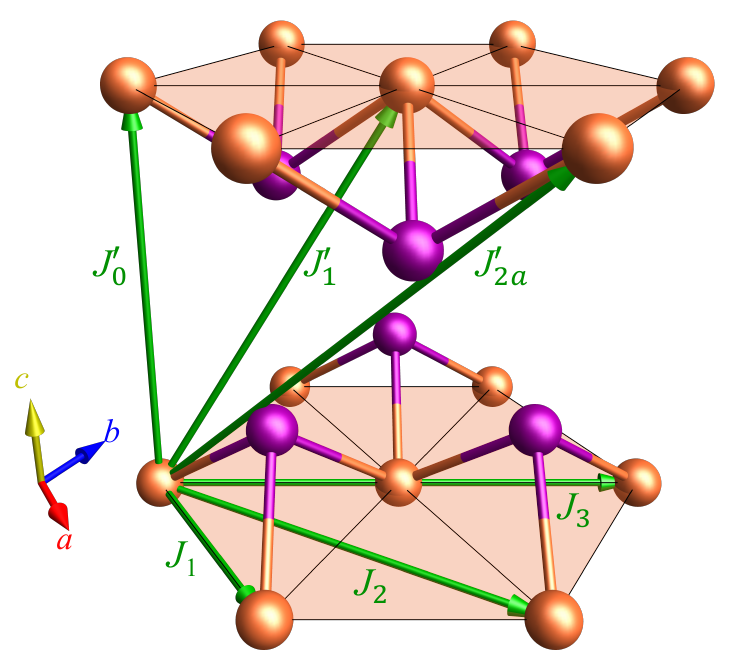}
}{
\caption{Crystal structure of FeI$_2$ and exchange pathways in the Hamiltonian model.}\label{fig:exchange}
}}\quad
{\capbtabbox[\FBwidth]{%
  \begin{tabular}{c|c|c||c|c|c|c|c||c}
\multicolumn{9}{c}{ } \\\hline\hline
\multicolumn{3}{c}{Nearest Neighbor} & \multicolumn{5}{|c|}{Further Neighbor} & Single-Ion \\ \hline\hline
   $J^{\pm}_1$ & $J^{\pm\pm}_1$ & $J^{z\pm}_1$ & $J^{\pm}_2$ & $J^{\pm}_3$ & $J^{'\pm}_0$ & $J^{'\pm}_1$ & $J^{'\pm}_{2a}$ & --   \\ \hline
   $-0.236$ & $-0.161$ & $-0.261$ & $0.026$ & $0.166$ & $0.037$ & $0.013$ & $0.068$ & --    \\ \hline\hline
	$J^{zz}_1$ & --  &  -- & $J^{zz}_2$ & $J^{zz}_3$ & $J^{\prime zz}_0$ & $J^{\prime zz}_1$ & $J^{\prime zz}_{2a}$ & $D$ \\ \hline
	$-0.236$ & -- & -- & $0.113$ & $0.211$ & $-0.036$ & $0.051$ & $0.073$ & $2.165$  \\ \hline \hline
\end{tabular}
}{%
  \caption{Hamiltonian Parameters of FeI$_2$. The values are in the unit of meV. }\label{tab:exchange}
}}
\end{floatrow}}
\end{figure}

\end{widetext}

\section{Experimental Methods}
\label{sec:exp-methods}
\begin{figure}[h!]
        \centering
        \includegraphics[width=0.9\columnwidth]{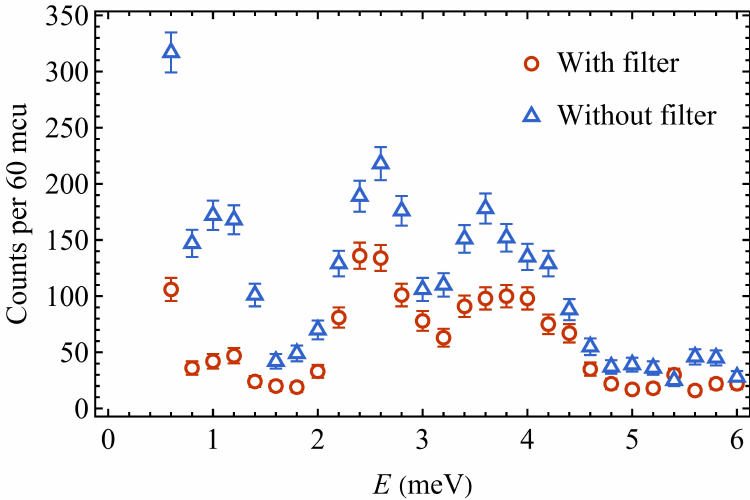}
        \caption{Comparison of INS spectra measured with and without a pyrolytic graphite filter. The spurious signals around 1~meV is strongly suppressed, confirming their origin from the scattering of higher-order neutrons. The efficiency of the filter is not optimal at $E_{\text{f}}\!=\!9.1$~meV which leads to residual spurious signals still being visible in the data.}
        \label{fig:AppxB}
\end{figure}

Single-crystal samples of $\sim$3\,gram FeI$_2$ were synthesized using the Bridgman method at the PARADIM facility, see Methods sections of Ref.\,\cite{Bai21, Bai23} for details. The inelastic neutron-scattering data presented in Fig.~\ref{fig:2} and \ref{fig:5} were respectively collected on the SEQUOIA time-of-flight spectrometer \cite{granroth2010,stone2014} at the Spallation Neutron Source (SNS) and the HB3 triple-axis spectrometer at High Flux Isotrope Reactor (HFIR) at Oak Ridge National Laboratory (ORNL), USA. SEQUOIA was operated with an incident neutron energy $E_{\text{i}}=12$\,meV and a high-resolution chopper mode yielding a full-width at half-maximum (FWHM) elastic energy resolution of $\Delta E \!=\! 0.27$\,meV. The sample was aligned in the $(h\,k\,0)$-plane. A liquid-helium cryostat was used to cool the sample mount to a base temperature of $T\!=\!1.8$\,K. Measurements at $T\!=\!2$\,K were performed by rotating the sample in steps of 0.5$^{\circ}$ for a coverage of 190$^{\circ}$. Data at alleviated temperatures $T\!=\!8,11,60$\,K were collected in steps of 1$^{\circ}$ rotation for a range of 200$^{\circ}$, 200$^{\circ}$ and 100$^{\circ}$, respectively. The counting time was $\sim2.5$\,mins per angle for all data.  Raw data were reduced in Mantid \cite{arnold2014} and subsequently processed and symmetrized in Horace \cite{ewings2016}. All 12 symmetry operations of the point group $\bar{3}m$ were used in the symmetrization process. 

HB3 was operated with pyrolytic graphite (PG 002) monochromator and analyzer set to a final neutron energy of $E_{\text{f}}\!=\!9.1$\,meV. Collimation of 48'-40'-40'-120' were selected, producing an elastic energy resolution FWHM of $\Delta E \approx0.47$\,meV. A close-cycle refrigerator reaching a base temperature of $T\!=\!4$\,K was employed, and data collected at $T\!=\!4,6,8,9,10,20,40,80$\,K. For either inelastic experiments, the neutron scattering intensity $\tilde{I}({\bf Q},E)=k_\text{i}/k_\text{f}[{\rm d^2 \sigma}/{\rm d} \Omega {\rm d} E_\text{f}]$ is plotted in arbitrary intensity units with ${\bf Q}=h\boldsymbol{a}^\ast + k\boldsymbol{b}^\ast + \ell\boldsymbol{c}^\ast$ projected in the recriprocal lattice of the hexagonal unit-cell. On HB3, the monitor count unit (mcu) of intensity corresponds to counting times of around 1 mcu $\approx$ 1 second for $E_\text{i}=E_\text{f}$ (elastic scattering).  A pyrolytic graphite filter was used to remove higher-order neutrons with $\lambda/2$  wavelength.  A direct comparison between spectra obtained with and without the PG filter is illustrated in Fig.~\ref{fig:AppxB}.

The elastic order parameter data presented in Fig.~\ref{fig:6} were measured at the HB3A DEMAND instrument \cite{cao2018} at HFIR. The instrument was operated in two-axis mode with an incoming neutron wavelength of {$\lambda_\text{i}=1.542$\AA} from a bent Si-220 monochromator and temperatures recorded down to $1.4$\,K in a cryomagnet.

\section{Derivation of the Quantum to Classical Crossover}
\label{sec:q2c-methods}

\subsection{Classical harmonic oscillator}

Consider the Hamiltonian of a classical harmonic oscillator:
\begin{equation}\nonumber
{\cal H} = \frac{{p}^2}{2m}  + \frac{m \omega_0^2}{2} x^2,
\end{equation}
At temperature $T$, the equipartion theorem tells us that $\langle {\cal H} \rangle = k_B T$ and 
\begin{eqnarray}\nonumber
 \left \langle \frac{m \omega_0^2}{2} x^2  \right \rangle   =  \left \langle \frac{{p}^2}{2m}  \right \rangle = \frac{k_B T}{2},
\end{eqnarray}
or 
\begin{equation}\nonumber
\langle x^2 \rangle =  \frac{k_B T}{m \omega_0^2}.
\end{equation}
We introduce now the two-point correlation function:
\begin{equation}\nonumber
C_{\rm cl}(t) =  \frac{1}{T} \int^{T}_{0} \langle x(t + \tau) x(\tau) \rangle d \tau
\end{equation}
where $T= 2\pi /\omega_0$ is the period of the harmonic oscillator. 
Since 
$$
x(t)= x(t=0) \cos(\omega_0 t )  + \frac{{\dot{x}} (0)}{\omega_0} \sin(\omega_0 t ),
$$
the Fourier transform ${\tilde C}_{\rm cl}(\omega)$ of the correlation function $C_{\rm cl}(t)$ has the form:
\begin{equation}\nonumber
{\tilde C}_{\rm cl}(\omega) = A [ \delta(\omega- \omega_0) + \delta(\omega + \omega_0)]
\end{equation}
The unknown constant $A$ is determined from the sum rule:
\begin{equation}\nonumber
\int d\omega {\tilde C}_{\rm cl}(\omega) = \langle x^2 \rangle =  \frac{k_B T}{m \omega_0^2} = 2 A,
\end{equation}
implying that
\begin{equation}
{\tilde C}_{\rm cl}(\omega) = \frac{k_B T}{2 m \omega_0^2} [ \delta(\omega- \omega_0) + \delta(\omega + \omega_0)]
\label{eq:cfclass}
\end{equation}

\subsection{Quantum Harmonic Oscillator}
Let us compute now the same correlation function for the quantum harmonic oscillator, whose Hamiltonian operator is:
\begin{equation}\nonumber
{\hat {\cal H}} = \frac{{\hat p}^2}{2m}  + \frac{m \omega_0^2{\hat x}^2}{2}
\end{equation}
This Hamiltonian operator can be diagonalized by introducing the creation and annihilation operators:
\begin{equation}
\begin{aligned}
\hat{a} & =\sqrt{\frac{m \omega_0}{2 \hbar}}\left(\hat{x}+\frac{i}{m \omega_0} \hat{p}\right) \\ \nonumber
\hat{a}^{\dagger} & =\sqrt{\frac{m \omega_0}{2 \hbar}}\left(\hat{x}-\frac{i}{m \omega_0} \hat{p}\right), \nonumber
\end{aligned}
\end{equation}
that lead to
\begin{equation}\nonumber
{\hat {\cal H}} = \hbar \omega_0 (\hat{a}^{\dagger}  \hat{a} + 1/2)
\end{equation}
The eigenstates of ${\hat {\cal H}}$ are also eigenstates of the number operator $\hat{n}= \hat{a}^{\dagger}  \hat{a}$ and $\hat{n} | n \rangle = n | n \rangle$
with $n$ being an integer larger or equal than zero:
\begin{equation}\nonumber
{\hat {\cal H}}  | n \rangle = \epsilon_n | n \rangle,
\end{equation}
with $\epsilon_n = \hbar \omega_0 (n+1/2)$.
The two-point correlation function
\begin{equation}\nonumber
C_{\rm Q} (t)  = \langle \hat{x}(t) \hat{x}(0) \rangle \equiv {\rm Tr}{\hat{\rho} \hat{x}(t) \hat{x}(0)}
\end{equation}
By using that $\hat{x}(t) = e^{i {\hat {\cal H}} t} \hat{x}(0)  e^{-i {\hat {\cal H}} t}$, we obtain
\begin{eqnarray}
C_{\rm Q} (t)  =  \sum_{n} \frac{e^{-\beta \epsilon_n}}{Z} \langle n | e^{i {\hat {\cal H}} t} (\hat{a}^{\dagger} + \hat{a}) e^{-i {\hat {\cal H}} t} (\hat{a}^{\dagger} + \hat{a}) | n \rangle  \nonumber
\end{eqnarray}
By inserting an identity,
\begin{eqnarray}
C_{\rm Q} (t)  = \frac{\hbar}{2 m \omega_0} \sum_{n, m} \frac{e^{-\beta \epsilon_n}}{Z} \langle n | e^{i {\hat {\cal H}} t} (\hat{a}^{\dagger} + \hat{a}) e^{-i {\hat {\cal H}} t} | m \rangle \langle m |(\hat{a}^{\dagger} + \hat{a}) | n \rangle \nonumber
\end{eqnarray}
and using the orthonormality condition $\langle n | m \rangle = \delta_{n,m}$, we obtain
\begin{equation}
C_{\rm Q} (t)  = \frac{\hbar}{2 m \omega_0}  [(1 + n_B(\omega_0)) e^{i \omega_0 t} + n_B(\omega_0) e^{-i \omega_0 t}], \nonumber
\end{equation}
which leads to the Fourier transform:
\begin{equation}
\tilde{C}_{\rm Q} (\omega)  = \frac{\hbar}{2 m \omega_0}  [(1 + n_B(\omega)) \delta(\omega- \omega_0) - (1+n_B(\omega)) \delta(\omega+\omega_0)],
\label{eq:cfquan}
\end{equation}
By comparing Eqs.~\eqref{eq:cfclass} and \eqref{eq:cfquan}, we obtain
\begin{equation}
\tilde{C}_{\rm Q} (\omega)  = \frac{\hbar \omega}{k_B T}  [1 + n_B(\omega)]  \tilde{C}_{\rm cl} (\omega).
\end{equation}
which is Eq.~\ref{eq:ho} from the main text.

\clearpage

\bibliographystyle{apsrev4-1}
\bibliography{refs}

\begin{thebibliography}{56}%
\makeatletter
\providecommand \@ifxundefined [1]{%
 \@ifx{#1\undefined}
}%
\providecommand \@ifnum [1]{%
 \ifnum #1\expandafter \@firstoftwo
 \else \expandafter \@secondoftwo
 \fi
}%
\providecommand \@ifx [1]{%
 \ifx #1\expandafter \@firstoftwo
 \else \expandafter \@secondoftwo
 \fi
}%
\providecommand \natexlab [1]{#1}%
\providecommand \enquote  [1]{``#1''}%
\providecommand \bibnamefont  [1]{#1}%
\providecommand \bibfnamefont [1]{#1}%
\providecommand \citenamefont [1]{#1}%
\providecommand \href@noop [0]{\@secondoftwo}%
\providecommand \href [0]{\begingroup \@sanitize@url \@href}%
\providecommand \@href[1]{\@@startlink{#1}\@@href}%
\providecommand \@@href[1]{\endgroup#1\@@endlink}%
\providecommand \@sanitize@url [0]{\catcode `\\12\catcode `\$12\catcode
  `\&12\catcode `\#12\catcode `\^12\catcode `\_12\catcode `\%12\relax}%
\providecommand \@@startlink[1]{}%
\providecommand \@@endlink[0]{}%
\providecommand \url  [0]{\begingroup\@sanitize@url \@url }%
\providecommand \@url [1]{\endgroup\@href {#1}{\urlprefix }}%
\providecommand \urlprefix  [0]{URL }%
\providecommand \Eprint [0]{\href }%
\providecommand \doibase [0]{http://dx.doi.org/}%
\providecommand \selectlanguage [0]{\@gobble}%
\providecommand \bibinfo  [0]{\@secondoftwo}%
\providecommand \bibfield  [0]{\@secondoftwo}%
\providecommand \translation [1]{[#1]}%
\providecommand \BibitemOpen [0]{}%
\providecommand \bibitemStop [0]{}%
\providecommand \bibitemNoStop [0]{.\EOS\space}%
\providecommand \EOS [0]{\spacefactor3000\relax}%
\providecommand \BibitemShut  [1]{\csname bibitem#1\endcsname}%
\let\auto@bib@innerbib\@empty
\bibitem [{\citenamefont {Marshall}\ and\ \citenamefont
  {Lowde}(1968)}]{marshall1_1968}%
  \BibitemOpen
  \bibfield  {author} {\bibinfo {author} {\bibfnamefont {W.}~\bibnamefont
  {Marshall}}\ and\ \bibinfo {author} {\bibfnamefont {R.}~\bibnamefont
  {Lowde}},\ }\href {\doibase 10.1088/0034-4885/31/2/305} {\bibfield  {journal}
  {\bibinfo  {journal} {Reports on Progress in Physics}\ }\textbf {\bibinfo
  {volume} {31}},\ \bibinfo {pages} {705} (\bibinfo {year} {1968})}\BibitemShut
  {NoStop}%
\bibitem [{\citenamefont {Lovesey}(1984)}]{lovesey_1984}%
  \BibitemOpen
  \bibfield  {author} {\bibinfo {author} {\bibfnamefont {S.~W.}\ \bibnamefont
  {Lovesey}},\ }\href {\doibase 10.1063/1.2815129} {\emph {\bibinfo {title}
  {Theory of neutron scattering from condensed matter}}}\ (\bibinfo
  {publisher} {Oxford University Press},\ \bibinfo {year} {1984})\BibitemShut
  {NoStop}%
\bibitem [{\citenamefont {Enderle}(2014)}]{enderle_2014}%
  \BibitemOpen
  \bibfield  {author} {\bibinfo {author} {\bibfnamefont {M.}~\bibnamefont
  {Enderle}},\ }\href {\doibase 10.1051/sfn/20141301002} {\bibfield  {journal}
  {\bibinfo  {journal} {{\'E}cole th{\'e}matique de la Soci{\'e}t{\'e}
  Fran{\c{c}}aise de la Neutronique}\ }\textbf {\bibinfo {volume} {13}},\
  \bibinfo {pages} {01002} (\bibinfo {year} {2014})}\BibitemShut {NoStop}%
\bibitem [{\citenamefont {Boothroyd}(2020)}]{boothroyd_2020}%
  \BibitemOpen
  \bibfield  {author} {\bibinfo {author} {\bibfnamefont {A.~T.}\ \bibnamefont
  {Boothroyd}},\ }\href {\doibase 10.1093/oso/9780198862314.001.0001} {\emph
  {\bibinfo {title} {Principles of neutron scattering from condensed matter}}}\
  (\bibinfo  {publisher} {Oxford University Press},\ \bibinfo {year}
  {2020})\BibitemShut {NoStop}%
\bibitem [{\citenamefont {Gagliano}\ and\ \citenamefont
  {Balseiro}(1988)}]{Gagliano88}%
  \BibitemOpen
  \bibfield  {author} {\bibinfo {author} {\bibfnamefont {E.~R.}\ \bibnamefont
  {Gagliano}}\ and\ \bibinfo {author} {\bibfnamefont {C.~A.}\ \bibnamefont
  {Balseiro}},\ }\href {\doibase 10.1103/PhysRevB.38.11766} {\bibfield
  {journal} {\bibinfo  {journal} {Phys. Rev. B}\ }\textbf {\bibinfo {volume}
  {38}},\ \bibinfo {pages} {11766} (\bibinfo {year} {1988})}\BibitemShut
  {NoStop}%
\bibitem [{\citenamefont {White}(1992)}]{White92}%
  \BibitemOpen
  \bibfield  {author} {\bibinfo {author} {\bibfnamefont {S.~R.}\ \bibnamefont
  {White}},\ }\href {\doibase 10.1103/PhysRevLett.69.2863} {\bibfield
  {journal} {\bibinfo  {journal} {Phys. Rev. Lett.}\ }\textbf {\bibinfo
  {volume} {69}},\ \bibinfo {pages} {2863} (\bibinfo {year}
  {1992})}\BibitemShut {NoStop}%
\bibitem [{\citenamefont {Schollw\"ock}(2005)}]{Schollwock05}%
  \BibitemOpen
  \bibfield  {author} {\bibinfo {author} {\bibfnamefont {U.}~\bibnamefont
  {Schollw\"ock}},\ }\href {\doibase 10.1103/RevModPhys.77.259} {\bibfield
  {journal} {\bibinfo  {journal} {Rev. Mod. Phys.}\ }\textbf {\bibinfo {volume}
  {77}},\ \bibinfo {pages} {259} (\bibinfo {year} {2005})}\BibitemShut
  {NoStop}%
\bibitem [{\citenamefont {Hallberg}(2006)}]{Hallberg06}%
  \BibitemOpen
  \bibfield  {author} {\bibinfo {author} {\bibfnamefont {K.~A.}\ \bibnamefont
  {Hallberg}},\ }\href {\doibase 10.1080/00018730600766432} {\bibfield
  {journal} {\bibinfo  {journal} {Advances in Physics}\ }\textbf {\bibinfo
  {volume} {55}},\ \bibinfo {pages} {477} (\bibinfo {year} {2006})}\BibitemShut
  {NoStop}%
\bibitem [{\citenamefont {Chi}\ \emph {et~al.}(2022)\citenamefont {Chi},
  \citenamefont {Liu}, \citenamefont {Wan}, \citenamefont {Liao},\ and\
  \citenamefont {Xiang}}]{Chi22}%
  \BibitemOpen
  \bibfield  {author} {\bibinfo {author} {\bibfnamefont {R.}~\bibnamefont
  {Chi}}, \bibinfo {author} {\bibfnamefont {Y.}~\bibnamefont {Liu}}, \bibinfo
  {author} {\bibfnamefont {Y.}~\bibnamefont {Wan}}, \bibinfo {author}
  {\bibfnamefont {H.-J.}\ \bibnamefont {Liao}}, \ and\ \bibinfo {author}
  {\bibfnamefont {T.}~\bibnamefont {Xiang}},\ }\href {\doibase
  10.1103/PhysRevLett.129.227201} {\bibfield  {journal} {\bibinfo  {journal}
  {Phys. Rev. Lett.}\ }\textbf {\bibinfo {volume} {129}},\ \bibinfo {pages}
  {227201} (\bibinfo {year} {2022})}\BibitemShut {NoStop}%
\bibitem [{\citenamefont {Hirsch}\ and\ \citenamefont
  {Schrieffer}(1983)}]{Hirsch83}%
  \BibitemOpen
  \bibfield  {author} {\bibinfo {author} {\bibfnamefont {J.~E.}\ \bibnamefont
  {Hirsch}}\ and\ \bibinfo {author} {\bibfnamefont {J.~R.}\ \bibnamefont
  {Schrieffer}},\ }\href {\doibase 10.1103/PhysRevB.28.5353} {\bibfield
  {journal} {\bibinfo  {journal} {Phys. Rev. B}\ }\textbf {\bibinfo {volume}
  {28}},\ \bibinfo {pages} {5353} (\bibinfo {year} {1983})}\BibitemShut
  {NoStop}%
\bibitem [{\citenamefont {Jarrell}\ and\ \citenamefont
  {Gubernatis}(1996)}]{JARRELL96}%
  \BibitemOpen
  \bibfield  {author} {\bibinfo {author} {\bibfnamefont {M.}~\bibnamefont
  {Jarrell}}\ and\ \bibinfo {author} {\bibfnamefont {J.}~\bibnamefont
  {Gubernatis}},\ }\href {\doibase
  https://doi.org/10.1016/0370-1573(95)00074-7} {\bibfield  {journal} {\bibinfo
   {journal} {Physics Reports}\ }\textbf {\bibinfo {volume} {269}},\ \bibinfo
  {pages} {133} (\bibinfo {year} {1996})}\BibitemShut {NoStop}%
\bibitem [{\citenamefont {Sandvik}(1998)}]{Sandvik98}%
  \BibitemOpen
  \bibfield  {author} {\bibinfo {author} {\bibfnamefont {A.~W.}\ \bibnamefont
  {Sandvik}},\ }\href {\doibase 10.1103/PhysRevB.57.10287} {\bibfield
  {journal} {\bibinfo  {journal} {Phys. Rev. B}\ }\textbf {\bibinfo {volume}
  {57}},\ \bibinfo {pages} {10287} (\bibinfo {year} {1998})}\BibitemShut
  {NoStop}%
\bibitem [{\citenamefont {Sandvik}(2016)}]{Sandvik16}%
  \BibitemOpen
  \bibfield  {author} {\bibinfo {author} {\bibfnamefont {A.~W.}\ \bibnamefont
  {Sandvik}},\ }\href {\doibase 10.1103/PhysRevE.94.063308} {\bibfield
  {journal} {\bibinfo  {journal} {Phys. Rev. E}\ }\textbf {\bibinfo {volume}
  {94}},\ \bibinfo {pages} {063308} (\bibinfo {year} {2016})}\BibitemShut
  {NoStop}%
\bibitem [{\citenamefont {Shao}\ and\ \citenamefont
  {Sandvik}(2023)}]{SHAO2023}%
  \BibitemOpen
  \bibfield  {author} {\bibinfo {author} {\bibfnamefont {H.}~\bibnamefont
  {Shao}}\ and\ \bibinfo {author} {\bibfnamefont {A.~W.}\ \bibnamefont
  {Sandvik}},\ }\href {\doibase https://doi.org/10.1016/j.physrep.2022.11.002}
  {\bibfield  {journal} {\bibinfo  {journal} {Physics Reports}\ }\textbf
  {\bibinfo {volume} {1003}},\ \bibinfo {pages} {1} (\bibinfo {year}
  {2023})}\BibitemShut {NoStop}%
\bibitem [{\citenamefont {Bai}\ \emph {et~al.}(2021)\citenamefont {Bai},
  \citenamefont {Zhang}, \citenamefont {Dun}, \citenamefont {Zhang},
  \citenamefont {Huang}, \citenamefont {Zhou}, \citenamefont {Stone},
  \citenamefont {Kolesnikov}, \citenamefont {Ye}, \citenamefont {Batista},\
  and\ \citenamefont {Mourigal}}]{Bai21}%
  \BibitemOpen
  \bibfield  {author} {\bibinfo {author} {\bibfnamefont {X.}~\bibnamefont
  {Bai}}, \bibinfo {author} {\bibfnamefont {S.-S.}\ \bibnamefont {Zhang}},
  \bibinfo {author} {\bibfnamefont {Z.}~\bibnamefont {Dun}}, \bibinfo {author}
  {\bibfnamefont {H.}~\bibnamefont {Zhang}}, \bibinfo {author} {\bibfnamefont
  {Q.}~\bibnamefont {Huang}}, \bibinfo {author} {\bibfnamefont
  {H.}~\bibnamefont {Zhou}}, \bibinfo {author} {\bibfnamefont {M.~B.}\
  \bibnamefont {Stone}}, \bibinfo {author} {\bibfnamefont {A.~I.}\ \bibnamefont
  {Kolesnikov}}, \bibinfo {author} {\bibfnamefont {F.}~\bibnamefont {Ye}},
  \bibinfo {author} {\bibfnamefont {C.~D.}\ \bibnamefont {Batista}}, \ and\
  \bibinfo {author} {\bibfnamefont {M.}~\bibnamefont {Mourigal}},\ }\href
  {\doibase 10.1038/s41567-020-01110-1} {\bibfield  {journal} {\bibinfo
  {journal} {Nature Physics}\ }\textbf {\bibinfo {volume} {17}},\ \bibinfo
  {pages} {467} (\bibinfo {year} {2021})}\BibitemShut {NoStop}%
\bibitem [{\citenamefont {Legros}\ \emph {et~al.}(2021)\citenamefont {Legros},
  \citenamefont {Zhang}, \citenamefont {Bai}, \citenamefont {Zhang},
  \citenamefont {Dun}, \citenamefont {Phelan}, \citenamefont {Batista},
  \citenamefont {Mourigal},\ and\ \citenamefont {Armitage}}]{Legros22}%
  \BibitemOpen
  \bibfield  {author} {\bibinfo {author} {\bibfnamefont {A.}~\bibnamefont
  {Legros}}, \bibinfo {author} {\bibfnamefont {S.-S.}\ \bibnamefont {Zhang}},
  \bibinfo {author} {\bibfnamefont {X.}~\bibnamefont {Bai}}, \bibinfo {author}
  {\bibfnamefont {H.}~\bibnamefont {Zhang}}, \bibinfo {author} {\bibfnamefont
  {Z.}~\bibnamefont {Dun}}, \bibinfo {author} {\bibfnamefont {W.~A.}\
  \bibnamefont {Phelan}}, \bibinfo {author} {\bibfnamefont {C.~D.}\
  \bibnamefont {Batista}}, \bibinfo {author} {\bibfnamefont {M.}~\bibnamefont
  {Mourigal}}, \ and\ \bibinfo {author} {\bibfnamefont {N.~P.}\ \bibnamefont
  {Armitage}},\ }\href {\doibase 10.1103/PhysRevLett.127.267201} {\bibfield
  {journal} {\bibinfo  {journal} {Phys. Rev. Lett.}\ }\textbf {\bibinfo
  {volume} {127}},\ \bibinfo {pages} {267201} (\bibinfo {year}
  {2021})}\BibitemShut {NoStop}%
\bibitem [{\citenamefont {Bai}\ \emph {et~al.}(2023)\citenamefont {Bai},
  \citenamefont {Zhang}, \citenamefont {Zhang}, \citenamefont {Dun},
  \citenamefont {Phelan}, \citenamefont {Garlea}, \citenamefont {Mourigal},\
  and\ \citenamefont {Batista}}]{Bai23}%
  \BibitemOpen
  \bibfield  {author} {\bibinfo {author} {\bibfnamefont {X.}~\bibnamefont
  {Bai}}, \bibinfo {author} {\bibfnamefont {S.-S.}\ \bibnamefont {Zhang}},
  \bibinfo {author} {\bibfnamefont {H.}~\bibnamefont {Zhang}}, \bibinfo
  {author} {\bibfnamefont {Z.}~\bibnamefont {Dun}}, \bibinfo {author}
  {\bibfnamefont {W.~A.}\ \bibnamefont {Phelan}}, \bibinfo {author}
  {\bibfnamefont {V.~O.}\ \bibnamefont {Garlea}}, \bibinfo {author}
  {\bibfnamefont {M.}~\bibnamefont {Mourigal}}, \ and\ \bibinfo {author}
  {\bibfnamefont {C.~D.}\ \bibnamefont {Batista}},\ }\href {\doibase
  10.1038/s41467-023-39940-1} {\bibfield  {journal} {\bibinfo  {journal}
  {Nature Communications}\ }\textbf {\bibinfo {volume} {14}},\ \bibinfo {pages}
  {4199} (\bibinfo {year} {2023})}\BibitemShut {NoStop}%
\bibitem [{\citenamefont {Zhang}\ and\ \citenamefont
  {Batista}(2021)}]{Zhang21}%
  \BibitemOpen
  \bibfield  {author} {\bibinfo {author} {\bibfnamefont {H.}~\bibnamefont
  {Zhang}}\ and\ \bibinfo {author} {\bibfnamefont {C.~D.}\ \bibnamefont
  {Batista}},\ }\href {\doibase 10.1103/PhysRevB.104.104409} {\bibfield
  {journal} {\bibinfo  {journal} {Phys. Rev. B}\ }\textbf {\bibinfo {volume}
  {104}},\ \bibinfo {pages} {104409} (\bibinfo {year} {2021})}\BibitemShut
  {NoStop}%
\bibitem [{\citenamefont {Dahlbom}\ \emph
  {et~al.}(2022{\natexlab{a}})\citenamefont {Dahlbom}, \citenamefont {Zhang},
  \citenamefont {Miles}, \citenamefont {Bai}, \citenamefont {Batista},\ and\
  \citenamefont {Barros}}]{Dahlbom22}%
  \BibitemOpen
  \bibfield  {author} {\bibinfo {author} {\bibfnamefont {D.}~\bibnamefont
  {Dahlbom}}, \bibinfo {author} {\bibfnamefont {H.}~\bibnamefont {Zhang}},
  \bibinfo {author} {\bibfnamefont {C.}~\bibnamefont {Miles}}, \bibinfo
  {author} {\bibfnamefont {X.}~\bibnamefont {Bai}}, \bibinfo {author}
  {\bibfnamefont {C.~D.}\ \bibnamefont {Batista}}, \ and\ \bibinfo {author}
  {\bibfnamefont {K.}~\bibnamefont {Barros}},\ }\href {\doibase
  10.1103/PhysRevB.106.054423} {\bibfield  {journal} {\bibinfo  {journal}
  {Phys. Rev. B}\ }\textbf {\bibinfo {volume} {106}},\ \bibinfo {pages}
  {054423} (\bibinfo {year} {2022}{\natexlab{a}})}\BibitemShut {NoStop}%
\bibitem [{\citenamefont {Dahlbom}\ \emph
  {et~al.}(2022{\natexlab{b}})\citenamefont {Dahlbom}, \citenamefont {Miles},
  \citenamefont {Zhang}, \citenamefont {Batista},\ and\ \citenamefont
  {Barros}}]{Dahlbom22b}%
  \BibitemOpen
  \bibfield  {author} {\bibinfo {author} {\bibfnamefont {D.}~\bibnamefont
  {Dahlbom}}, \bibinfo {author} {\bibfnamefont {C.}~\bibnamefont {Miles}},
  \bibinfo {author} {\bibfnamefont {H.}~\bibnamefont {Zhang}}, \bibinfo
  {author} {\bibfnamefont {C.~D.}\ \bibnamefont {Batista}}, \ and\ \bibinfo
  {author} {\bibfnamefont {K.}~\bibnamefont {Barros}},\ }\href {\doibase
  10.1103/PhysRevB.106.235154} {\bibfield  {journal} {\bibinfo  {journal}
  {Phys. Rev. B}\ }\textbf {\bibinfo {volume} {106}},\ \bibinfo {pages}
  {235154} (\bibinfo {year} {2022}{\natexlab{b}})}\BibitemShut {NoStop}%
\bibitem [{\citenamefont {Schofield}(1960)}]{Schofield_1960}%
  \BibitemOpen
  \bibfield  {author} {\bibinfo {author} {\bibfnamefont {P.}~\bibnamefont
  {Schofield}},\ }\href {\doibase 10.1103/PhysRevLett.4.239} {\bibfield
  {journal} {\bibinfo  {journal} {Phys. Rev. Lett.}\ }\textbf {\bibinfo
  {volume} {4}},\ \bibinfo {pages} {239} (\bibinfo {year} {1960})}\BibitemShut
  {NoStop}%
\bibitem [{\citenamefont {Zhang}\ \emph {et~al.}(2019)\citenamefont {Zhang},
  \citenamefont {Changlani}, \citenamefont {Plumb}, \citenamefont
  {Tchernyshyov},\ and\ \citenamefont {Moessner}}]{Zhang_2019}%
  \BibitemOpen
  \bibfield  {author} {\bibinfo {author} {\bibfnamefont {S.}~\bibnamefont
  {Zhang}}, \bibinfo {author} {\bibfnamefont {H.~J.}\ \bibnamefont
  {Changlani}}, \bibinfo {author} {\bibfnamefont {K.~W.}\ \bibnamefont
  {Plumb}}, \bibinfo {author} {\bibfnamefont {O.}~\bibnamefont {Tchernyshyov}},
  \ and\ \bibinfo {author} {\bibfnamefont {R.}~\bibnamefont {Moessner}},\
  }\href {\doibase 10.1103/PhysRevLett.122.167203} {\bibfield  {journal}
  {\bibinfo  {journal} {Phys. Rev. Lett.}\ }\textbf {\bibinfo {volume} {122}},\
  \bibinfo {pages} {167203} (\bibinfo {year} {2019})}\BibitemShut {NoStop}%
\bibitem [{\citenamefont {Plumb}\ \emph {et~al.}(2019)\citenamefont {Plumb},
  \citenamefont {Changlani}, \citenamefont {Scheie}, \citenamefont {Zhang},
  \citenamefont {Krizan}, \citenamefont {{Rodriguez-Rivera}}, \citenamefont
  {Qiu}, \citenamefont {Winn}, \citenamefont {Cava},\ and\ \citenamefont
  {Broholm}}]{Plumb19}%
  \BibitemOpen
  \bibfield  {author} {\bibinfo {author} {\bibfnamefont {K.~W.}\ \bibnamefont
  {Plumb}}, \bibinfo {author} {\bibfnamefont {H.~J.}\ \bibnamefont
  {Changlani}}, \bibinfo {author} {\bibfnamefont {A.}~\bibnamefont {Scheie}},
  \bibinfo {author} {\bibfnamefont {S.}~\bibnamefont {Zhang}}, \bibinfo
  {author} {\bibfnamefont {J.~W.}\ \bibnamefont {Krizan}}, \bibinfo {author}
  {\bibfnamefont {J.~A.}\ \bibnamefont {{Rodriguez-Rivera}}}, \bibinfo {author}
  {\bibfnamefont {Y.}~\bibnamefont {Qiu}}, \bibinfo {author} {\bibfnamefont
  {B.}~\bibnamefont {Winn}}, \bibinfo {author} {\bibfnamefont {R.~J.}\
  \bibnamefont {Cava}}, \ and\ \bibinfo {author} {\bibfnamefont {C.~L.}\
  \bibnamefont {Broholm}},\ }\href {\doibase 10.1038/s41567-018-0317-3}
  {\bibfield  {journal} {\bibinfo  {journal} {Nature Physics}\ }\textbf
  {\bibinfo {volume} {15}},\ \bibinfo {pages} {54} (\bibinfo {year}
  {2019})}\BibitemShut {NoStop}%
\bibitem [{\citenamefont {Bai}\ \emph {et~al.}(2019)\citenamefont {Bai},
  \citenamefont {Paddison}, \citenamefont {Kapit}, \citenamefont {Koohpayeh},
  \citenamefont {Wen}, \citenamefont {Dutton}, \citenamefont {Savici},
  \citenamefont {Kolesnikov}, \citenamefont {Granroth}, \citenamefont
  {Broholm}, \citenamefont {Chalker},\ and\ \citenamefont {Mourigal}}]{Bai19}%
  \BibitemOpen
  \bibfield  {author} {\bibinfo {author} {\bibfnamefont {X.}~\bibnamefont
  {Bai}}, \bibinfo {author} {\bibfnamefont {J.~A.~M.}\ \bibnamefont
  {Paddison}}, \bibinfo {author} {\bibfnamefont {E.}~\bibnamefont {Kapit}},
  \bibinfo {author} {\bibfnamefont {S.~M.}\ \bibnamefont {Koohpayeh}}, \bibinfo
  {author} {\bibfnamefont {J.-J.}\ \bibnamefont {Wen}}, \bibinfo {author}
  {\bibfnamefont {S.~E.}\ \bibnamefont {Dutton}}, \bibinfo {author}
  {\bibfnamefont {A.~T.}\ \bibnamefont {Savici}}, \bibinfo {author}
  {\bibfnamefont {A.~I.}\ \bibnamefont {Kolesnikov}}, \bibinfo {author}
  {\bibfnamefont {G.~E.}\ \bibnamefont {Granroth}}, \bibinfo {author}
  {\bibfnamefont {C.~L.}\ \bibnamefont {Broholm}}, \bibinfo {author}
  {\bibfnamefont {J.~T.}\ \bibnamefont {Chalker}}, \ and\ \bibinfo {author}
  {\bibfnamefont {M.}~\bibnamefont {Mourigal}},\ }\href {\doibase
  10.1103/PhysRevLett.122.097201} {\bibfield  {journal} {\bibinfo  {journal}
  {Physical Review Letters}\ }\textbf {\bibinfo {volume} {122}},\ \bibinfo
  {pages} {097201} (\bibinfo {year} {2019})}\BibitemShut {NoStop}%
\bibitem [{\citenamefont {Remund}\ \emph {et~al.}(2022)\citenamefont {Remund},
  \citenamefont {Pohle}, \citenamefont {Akagi}, \citenamefont {Romh\'anyi},\
  and\ \citenamefont {Shannon}}]{Remund22}%
  \BibitemOpen
  \bibfield  {author} {\bibinfo {author} {\bibfnamefont {K.}~\bibnamefont
  {Remund}}, \bibinfo {author} {\bibfnamefont {R.}~\bibnamefont {Pohle}},
  \bibinfo {author} {\bibfnamefont {Y.}~\bibnamefont {Akagi}}, \bibinfo
  {author} {\bibfnamefont {J.}~\bibnamefont {Romh\'anyi}}, \ and\ \bibinfo
  {author} {\bibfnamefont {N.}~\bibnamefont {Shannon}},\ }\href {\doibase
  10.1103/PhysRevResearch.4.033106} {\bibfield  {journal} {\bibinfo  {journal}
  {Phys. Rev. Res.}\ }\textbf {\bibinfo {volume} {4}},\ \bibinfo {pages}
  {033106} (\bibinfo {year} {2022})}\BibitemShut {NoStop}%
\bibitem [{\citenamefont {Pohle}\ \emph {et~al.}(2023)\citenamefont {Pohle},
  \citenamefont {Shannon},\ and\ \citenamefont {Motome}}]{Pohle23}%
  \BibitemOpen
  \bibfield  {author} {\bibinfo {author} {\bibfnamefont {R.}~\bibnamefont
  {Pohle}}, \bibinfo {author} {\bibfnamefont {N.}~\bibnamefont {Shannon}}, \
  and\ \bibinfo {author} {\bibfnamefont {Y.}~\bibnamefont {Motome}},\ }\href
  {\doibase 10.1103/PhysRevB.107.L140403} {\bibfield  {journal} {\bibinfo
  {journal} {Phys. Rev. B}\ }\textbf {\bibinfo {volume} {107}},\ \bibinfo
  {pages} {L140403} (\bibinfo {year} {2023})}\BibitemShut {NoStop}%
\bibitem [{\citenamefont {Do}\ \emph {et~al.}(2023)\citenamefont {Do},
  \citenamefont {Zhang}, \citenamefont {Dahlbom}, \citenamefont {Williams},
  \citenamefont {Garlea}, \citenamefont {Hong}, \citenamefont {Jang},
  \citenamefont {Cheong}, \citenamefont {Park}, \citenamefont {Barros},
  \citenamefont {Batista},\ and\ \citenamefont {Christianson}}]{Hwan23}%
  \BibitemOpen
  \bibfield  {author} {\bibinfo {author} {\bibfnamefont {S.-H.}\ \bibnamefont
  {Do}}, \bibinfo {author} {\bibfnamefont {H.}~\bibnamefont {Zhang}}, \bibinfo
  {author} {\bibfnamefont {D.~A.}\ \bibnamefont {Dahlbom}}, \bibinfo {author}
  {\bibfnamefont {T.~J.}\ \bibnamefont {Williams}}, \bibinfo {author}
  {\bibfnamefont {V.~O.}\ \bibnamefont {Garlea}}, \bibinfo {author}
  {\bibfnamefont {T.}~\bibnamefont {Hong}}, \bibinfo {author} {\bibfnamefont
  {T.-H.}\ \bibnamefont {Jang}}, \bibinfo {author} {\bibfnamefont {S.-W.}\
  \bibnamefont {Cheong}}, \bibinfo {author} {\bibfnamefont {J.-H.}\
  \bibnamefont {Park}}, \bibinfo {author} {\bibfnamefont {K.}~\bibnamefont
  {Barros}}, \bibinfo {author} {\bibfnamefont {C.~D.}\ \bibnamefont {Batista}},
  \ and\ \bibinfo {author} {\bibfnamefont {A.~D.}\ \bibnamefont
  {Christianson}},\ }\href {\doibase 10.1038/s41535-022-00526-7} {\bibfield
  {journal} {\bibinfo  {journal} {npj Quantum Materials}\ }\textbf {\bibinfo
  {volume} {8}},\ \bibinfo {pages} {5} (\bibinfo {year} {2023})}\BibitemShut
  {NoStop}%
\bibitem [{\citenamefont {Huberman}\ \emph {et~al.}(2008)\citenamefont
  {Huberman}, \citenamefont {Tennant}, \citenamefont {Cowley}, \citenamefont
  {Coldea},\ and\ \citenamefont {Frost}}]{Huberman08}%
  \BibitemOpen
  \bibfield  {author} {\bibinfo {author} {\bibfnamefont {T.}~\bibnamefont
  {Huberman}}, \bibinfo {author} {\bibfnamefont {D.~A.}\ \bibnamefont
  {Tennant}}, \bibinfo {author} {\bibfnamefont {R.~A.}\ \bibnamefont {Cowley}},
  \bibinfo {author} {\bibfnamefont {R.}~\bibnamefont {Coldea}}, \ and\ \bibinfo
  {author} {\bibfnamefont {C.~D.}\ \bibnamefont {Frost}},\ }\href {\doibase
  10.1088/1742-5468/2008/05/P05017} {\bibfield  {journal} {\bibinfo  {journal}
  {Journal of Statistical Mechanics: Theory and Experiment}\ }\textbf {\bibinfo
  {volume} {2008}},\ \bibinfo {pages} {P05017} (\bibinfo {year}
  {2008})}\BibitemShut {NoStop}%
\bibitem [{\citenamefont {McGuire}\ \emph {et~al.}(2017)\citenamefont
  {McGuire}, \citenamefont {{McGuire}},\ and\ \citenamefont {A.}}]{McGuire17}%
  \BibitemOpen
  \bibfield  {author} {\bibinfo {author} {\bibfnamefont {M.}~\bibnamefont
  {McGuire}}, \bibinfo {author} {\bibnamefont {{McGuire}}}, \ and\ \bibinfo
  {author} {\bibfnamefont {M.}~\bibnamefont {A.}},\ }\href {\doibase
  10.3390/cryst7050121} {\bibfield  {journal} {\bibinfo  {journal} {Crystals}\
  }\textbf {\bibinfo {volume} {7}},\ \bibinfo {pages} {121} (\bibinfo {year}
  {2017})}\BibitemShut {NoStop}%
\bibitem [{\citenamefont {Bertrand}\ \emph {et~al.}(1974)\citenamefont
  {Bertrand}, \citenamefont {Fert},\ and\ \citenamefont
  {Gelard}}]{bertrand1974susceptibilite}%
  \BibitemOpen
  \bibfield  {author} {\bibinfo {author} {\bibfnamefont {Y.}~\bibnamefont
  {Bertrand}}, \bibinfo {author} {\bibfnamefont {A.}~\bibnamefont {Fert}}, \
  and\ \bibinfo {author} {\bibfnamefont {J.}~\bibnamefont {Gelard}},\ }\href
  {\doibase 10.1051/jphys:01974003504038500} {\bibfield  {journal} {\bibinfo
  {journal} {Journal de Physique}\ }\textbf {\bibinfo {volume} {35}},\ \bibinfo
  {pages} {385} (\bibinfo {year} {1974})}\BibitemShut {NoStop}%
\bibitem [{\citenamefont {Gelard}\ \emph {et~al.}(1974)\citenamefont {Gelard},
  \citenamefont {Fert}, \citenamefont {Meriel},\ and\ \citenamefont
  {Allain}}]{gelard1974magnetic}%
  \BibitemOpen
  \bibfield  {author} {\bibinfo {author} {\bibfnamefont {J.}~\bibnamefont
  {Gelard}}, \bibinfo {author} {\bibfnamefont {A.}~\bibnamefont {Fert}},
  \bibinfo {author} {\bibfnamefont {P.}~\bibnamefont {Meriel}}, \ and\ \bibinfo
  {author} {\bibfnamefont {Y.}~\bibnamefont {Allain}},\ }\href {\doibase
  10.1016/0038-1098(74)90213-0} {\bibfield  {journal} {\bibinfo  {journal}
  {Solid State Communications}\ }\textbf {\bibinfo {volume} {14}},\ \bibinfo
  {pages} {187} (\bibinfo {year} {1974})}\BibitemShut {NoStop}%
\bibitem [{\citenamefont {Fert}\ \emph {et~al.}(1973)\citenamefont {Fert},
  \citenamefont {Gelard},\ and\ \citenamefont {Carrara}}]{fert1973phase}%
  \BibitemOpen
  \bibfield  {author} {\bibinfo {author} {\bibfnamefont {A.}~\bibnamefont
  {Fert}}, \bibinfo {author} {\bibfnamefont {J.}~\bibnamefont {Gelard}}, \ and\
  \bibinfo {author} {\bibfnamefont {P.}~\bibnamefont {Carrara}},\ }\href
  {\doibase 10.1016/0038-1098(73)90568-1} {\bibfield  {journal} {\bibinfo
  {journal} {Solid State Communications}\ }\textbf {\bibinfo {volume} {13}},\
  \bibinfo {pages} {1219} (\bibinfo {year} {1973})}\BibitemShut {NoStop}%
\bibitem [{\citenamefont {Wiedenmann}\ \emph {et~al.}(1988)\citenamefont
  {Wiedenmann}, \citenamefont {Regnault}, \citenamefont {Burlet}, \citenamefont
  {Rossat-Mignod}, \citenamefont {Kound{\'e}},\ and\ \citenamefont
  {Billerey}}]{wiedenmann1988neutron}%
  \BibitemOpen
  \bibfield  {author} {\bibinfo {author} {\bibfnamefont {A.}~\bibnamefont
  {Wiedenmann}}, \bibinfo {author} {\bibfnamefont {L.}~\bibnamefont
  {Regnault}}, \bibinfo {author} {\bibfnamefont {P.}~\bibnamefont {Burlet}},
  \bibinfo {author} {\bibfnamefont {J.}~\bibnamefont {Rossat-Mignod}}, \bibinfo
  {author} {\bibfnamefont {O.}~\bibnamefont {Kound{\'e}}}, \ and\ \bibinfo
  {author} {\bibfnamefont {D.}~\bibnamefont {Billerey}},\ }\href {\doibase
  10.1016/0304-8853(88)90143-6} {\bibfield  {journal} {\bibinfo  {journal}
  {Journal of magnetism and magnetic materials}\ }\textbf {\bibinfo {volume}
  {74}},\ \bibinfo {pages} {7} (\bibinfo {year} {1988})}\BibitemShut {NoStop}%
\bibitem [{\citenamefont {Katsumata}\ \emph {et~al.}(2010)\citenamefont
  {Katsumata}, \citenamefont {Katori}, \citenamefont {Kimura}, \citenamefont
  {Narumi}, \citenamefont {Hagiwara},\ and\ \citenamefont
  {Kindo}}]{katsumata2010phase}%
  \BibitemOpen
  \bibfield  {author} {\bibinfo {author} {\bibfnamefont {K.}~\bibnamefont
  {Katsumata}}, \bibinfo {author} {\bibfnamefont {H.~A.}\ \bibnamefont
  {Katori}}, \bibinfo {author} {\bibfnamefont {S.}~\bibnamefont {Kimura}},
  \bibinfo {author} {\bibfnamefont {Y.}~\bibnamefont {Narumi}}, \bibinfo
  {author} {\bibfnamefont {M.}~\bibnamefont {Hagiwara}}, \ and\ \bibinfo
  {author} {\bibfnamefont {K.}~\bibnamefont {Kindo}},\ }\href@noop {}
  {\bibfield  {journal} {\bibinfo  {journal} {Physical Review B}\ }\textbf
  {\bibinfo {volume} {82}},\ \bibinfo {pages} {104402} (\bibinfo {year}
  {2010})}\BibitemShut {NoStop}%
\bibitem [{\citenamefont {Petitgrand}\ and\ \citenamefont
  {Meyer}(1976)}]{petitgrand1976far}%
  \BibitemOpen
  \bibfield  {author} {\bibinfo {author} {\bibfnamefont {D.}~\bibnamefont
  {Petitgrand}}\ and\ \bibinfo {author} {\bibfnamefont {P.}~\bibnamefont
  {Meyer}},\ }\href {\doibase 10.1051/jphys:0197600370120141700} {\bibfield
  {journal} {\bibinfo  {journal} {Journal de Physique}\ }\textbf {\bibinfo
  {volume} {37}},\ \bibinfo {pages} {1417} (\bibinfo {year}
  {1976})}\BibitemShut {NoStop}%
\bibitem [{\citenamefont {Fert}\ \emph {et~al.}(1978)\citenamefont {Fert},
  \citenamefont {Bertrand}, \citenamefont {Leotin}, \citenamefont {Ousset},
  \citenamefont {Magari{\~n}o},\ and\ \citenamefont
  {Tuchendler}}]{fert1978excitation}%
  \BibitemOpen
  \bibfield  {author} {\bibinfo {author} {\bibfnamefont {A.}~\bibnamefont
  {Fert}}, \bibinfo {author} {\bibfnamefont {D.}~\bibnamefont {Bertrand}},
  \bibinfo {author} {\bibfnamefont {J.}~\bibnamefont {Leotin}}, \bibinfo
  {author} {\bibfnamefont {J.}~\bibnamefont {Ousset}}, \bibinfo {author}
  {\bibfnamefont {J.}~\bibnamefont {Magari{\~n}o}}, \ and\ \bibinfo {author}
  {\bibfnamefont {J.}~\bibnamefont {Tuchendler}},\ }\href {\doibase
  10.1016/0038-1098(78)90721-4} {\bibfield  {journal} {\bibinfo  {journal}
  {Solid State Communications}\ }\textbf {\bibinfo {volume} {26}},\ \bibinfo
  {pages} {693} (\bibinfo {year} {1978})}\BibitemShut {NoStop}%
\bibitem [{\citenamefont {Petitgrand}\ \emph {et~al.}(1980)\citenamefont
  {Petitgrand}, \citenamefont {Brun},\ and\ \citenamefont
  {Meyer}}]{petitgrand1980magnetic}%
  \BibitemOpen
  \bibfield  {author} {\bibinfo {author} {\bibfnamefont {D.}~\bibnamefont
  {Petitgrand}}, \bibinfo {author} {\bibfnamefont {A.}~\bibnamefont {Brun}}, \
  and\ \bibinfo {author} {\bibfnamefont {P.}~\bibnamefont {Meyer}},\ }\href
  {\doibase 10.1016/0304-8853(80)91097-5} {\bibfield  {journal} {\bibinfo
  {journal} {Journal of Magnetism and Magnetic Materials}\ }\textbf {\bibinfo
  {volume} {15}},\ \bibinfo {pages} {381} (\bibinfo {year} {1980})}\BibitemShut
  {NoStop}%
\bibitem [{\citenamefont {Lockwood}\ \emph {et~al.}(1994)\citenamefont
  {Lockwood}, \citenamefont {Mischler},\ and\ \citenamefont
  {Zwick}}]{lockwood1994raman}%
  \BibitemOpen
  \bibfield  {author} {\bibinfo {author} {\bibfnamefont {D.}~\bibnamefont
  {Lockwood}}, \bibinfo {author} {\bibfnamefont {G.}~\bibnamefont {Mischler}},
  \ and\ \bibinfo {author} {\bibfnamefont {A.}~\bibnamefont {Zwick}},\ }\href
  {\doibase 10.1088/0953-8984/6/32/013} {\bibfield  {journal} {\bibinfo
  {journal} {Journal of Physics: Condensed Matter}\ }\textbf {\bibinfo {volume}
  {6}},\ \bibinfo {pages} {6515} (\bibinfo {year} {1994})}\BibitemShut
  {NoStop}%
\bibitem [{\citenamefont {Katsumata}\ \emph
  {et~al.}(2000{\natexlab{a}})\citenamefont {Katsumata}, \citenamefont
  {Yamaguchi}, \citenamefont {Hagiwara}, \citenamefont {Tokunaga},
  \citenamefont {Mikeska}, \citenamefont {Goy},\ and\ \citenamefont
  {Gross}}]{katsumata2000single}%
  \BibitemOpen
  \bibfield  {author} {\bibinfo {author} {\bibfnamefont {K.}~\bibnamefont
  {Katsumata}}, \bibinfo {author} {\bibfnamefont {H.}~\bibnamefont
  {Yamaguchi}}, \bibinfo {author} {\bibfnamefont {M.}~\bibnamefont {Hagiwara}},
  \bibinfo {author} {\bibfnamefont {M.}~\bibnamefont {Tokunaga}}, \bibinfo
  {author} {\bibfnamefont {H.-J.}\ \bibnamefont {Mikeska}}, \bibinfo {author}
  {\bibfnamefont {P.}~\bibnamefont {Goy}}, \ and\ \bibinfo {author}
  {\bibfnamefont {M.}~\bibnamefont {Gross}},\ }\href {\doibase
  10.1103/PhysRevB.61.11632} {\bibfield  {journal} {\bibinfo  {journal}
  {Physical Review B}\ }\textbf {\bibinfo {volume} {61}},\ \bibinfo {pages}
  {11632} (\bibinfo {year} {2000}{\natexlab{a}})}\BibitemShut {NoStop}%
\bibitem [{\citenamefont {Katsumata}\ \emph
  {et~al.}(2000{\natexlab{b}})\citenamefont {Katsumata}, \citenamefont
  {Hagiwara}, \citenamefont {Tokunaga},\ and\ \citenamefont
  {Yamaguchi}}]{katsumata2000observation}%
  \BibitemOpen
  \bibfield  {author} {\bibinfo {author} {\bibfnamefont {K.}~\bibnamefont
  {Katsumata}}, \bibinfo {author} {\bibfnamefont {M.}~\bibnamefont {Hagiwara}},
  \bibinfo {author} {\bibfnamefont {M.}~\bibnamefont {Tokunaga}}, \ and\
  \bibinfo {author} {\bibfnamefont {H.}~\bibnamefont {Yamaguchi}},\ }\href
  {\doibase 10.1063/1.373256} {\bibfield  {journal} {\bibinfo  {journal}
  {Journal of Applied Physics}\ }\textbf {\bibinfo {volume} {87}},\ \bibinfo
  {pages} {5085} (\bibinfo {year} {2000}{\natexlab{b}})}\BibitemShut {NoStop}%
\bibitem [{\citenamefont {Petitgrand}\ \emph {et~al.}(1979)\citenamefont
  {Petitgrand}, \citenamefont {Hennion},\ and\ \citenamefont
  {Escribe}}]{petitgrand1979neutron}%
  \BibitemOpen
  \bibfield  {author} {\bibinfo {author} {\bibfnamefont {D.}~\bibnamefont
  {Petitgrand}}, \bibinfo {author} {\bibfnamefont {B.}~\bibnamefont {Hennion}},
  \ and\ \bibinfo {author} {\bibfnamefont {C.}~\bibnamefont {Escribe}},\ }\href
  {\doibase 10.1016/0304-8853(79)90138-0} {\bibfield  {journal} {\bibinfo
  {journal} {Journal of Magnetism and Magnetic Materials}\ }\textbf {\bibinfo
  {volume} {14}},\ \bibinfo {pages} {275} (\bibinfo {year} {1979})}\BibitemShut
  {NoStop}%
\bibitem [{\citenamefont {Silberglitt}\ and\ \citenamefont
  {Torrance~Jr}(1970)}]{silberglitt1970effect}%
  \BibitemOpen
  \bibfield  {author} {\bibinfo {author} {\bibfnamefont {R.}~\bibnamefont
  {Silberglitt}}\ and\ \bibinfo {author} {\bibfnamefont {J.~B.}\ \bibnamefont
  {Torrance~Jr}},\ }\href {\doibase 10.1103/PhysRevB.2.772} {\bibfield
  {journal} {\bibinfo  {journal} {Physical Review B}\ }\textbf {\bibinfo
  {volume} {2}},\ \bibinfo {pages} {772} (\bibinfo {year} {1970})}\BibitemShut
  {NoStop}%
\bibitem [{\citenamefont {Ono}\ \emph {et~al.}(1971)\citenamefont {Ono},
  \citenamefont {Mikado},\ and\ \citenamefont {Oguchi}}]{ono1971two}%
  \BibitemOpen
  \bibfield  {author} {\bibinfo {author} {\bibfnamefont {I.}~\bibnamefont
  {Ono}}, \bibinfo {author} {\bibfnamefont {S.}~\bibnamefont {Mikado}}, \ and\
  \bibinfo {author} {\bibfnamefont {T.}~\bibnamefont {Oguchi}},\ }\href
  {\doibase 10.1143/JPSJ.30.358} {\bibfield  {journal} {\bibinfo  {journal}
  {Journal of the Physical Society of Japan}\ }\textbf {\bibinfo {volume}
  {30}},\ \bibinfo {pages} {358} (\bibinfo {year} {1971})}\BibitemShut
  {NoStop}%
\bibitem [{\citenamefont {Savin}\ \emph {et~al.}(2012)\citenamefont {Savin},
  \citenamefont {Kosevich},\ and\ \citenamefont {Cantarero}}]{Savin12}%
  \BibitemOpen
  \bibfield  {author} {\bibinfo {author} {\bibfnamefont {A.~V.}\ \bibnamefont
  {Savin}}, \bibinfo {author} {\bibfnamefont {Y.~A.}\ \bibnamefont {Kosevich}},
  \ and\ \bibinfo {author} {\bibfnamefont {A.}~\bibnamefont {Cantarero}},\
  }\href {\doibase 10.1103/PhysRevB.86.064305} {\bibfield  {journal} {\bibinfo
  {journal} {Phys. Rev. B}\ }\textbf {\bibinfo {volume} {86}},\ \bibinfo
  {pages} {064305} (\bibinfo {year} {2012})}\BibitemShut {NoStop}%
\bibitem [{\citenamefont {Dahlbom}\ \emph {et~al.}(2023)\citenamefont
  {Dahlbom}, \citenamefont {Zhang}, \citenamefont {Laraib}, \citenamefont
  {Pajerowski}, \citenamefont {Barros},\ and\ \citenamefont
  {Batista}}]{Dahlbom23}%
  \BibitemOpen
  \bibfield  {author} {\bibinfo {author} {\bibfnamefont {D.}~\bibnamefont
  {Dahlbom}}, \bibinfo {author} {\bibfnamefont {H.}~\bibnamefont {Zhang}},
  \bibinfo {author} {\bibfnamefont {Z.}~\bibnamefont {Laraib}}, \bibinfo
  {author} {\bibfnamefont {D.~M.}\ \bibnamefont {Pajerowski}}, \bibinfo
  {author} {\bibfnamefont {K.}~\bibnamefont {Barros}}, \ and\ \bibinfo {author}
  {\bibfnamefont {C.}~\bibnamefont {Batista}},\ }\href {\doibase
  10.48550/arXiv.2304.03874} {\bibfield  {journal} {\bibinfo  {journal} {arXiv
  preprint arXiv:2304.03874}\ } (\bibinfo {year} {2023}),\
  10.48550/arXiv.2304.03874}\BibitemShut {NoStop}%
\bibitem [{\citenamefont {Samarakoon}\ \emph {et~al.}(2017)\citenamefont
  {Samarakoon}, \citenamefont {Banerjee}, \citenamefont {Zhang}, \citenamefont
  {Kamiya}, \citenamefont {Nagler}, \citenamefont {Tennant}, \citenamefont
  {Lee},\ and\ \citenamefont {Batista}}]{Samarakoon17}%
  \BibitemOpen
  \bibfield  {author} {\bibinfo {author} {\bibfnamefont {A.~M.}\ \bibnamefont
  {Samarakoon}}, \bibinfo {author} {\bibfnamefont {A.}~\bibnamefont
  {Banerjee}}, \bibinfo {author} {\bibfnamefont {S.-S.}\ \bibnamefont {Zhang}},
  \bibinfo {author} {\bibfnamefont {Y.}~\bibnamefont {Kamiya}}, \bibinfo
  {author} {\bibfnamefont {S.~E.}\ \bibnamefont {Nagler}}, \bibinfo {author}
  {\bibfnamefont {D.~A.}\ \bibnamefont {Tennant}}, \bibinfo {author}
  {\bibfnamefont {S.-H.}\ \bibnamefont {Lee}}, \ and\ \bibinfo {author}
  {\bibfnamefont {C.~D.}\ \bibnamefont {Batista}},\ }\href {\doibase
  10.1103/PhysRevB.96.134408} {\bibfield  {journal} {\bibinfo  {journal} {Phys.
  Rev. B}\ }\textbf {\bibinfo {volume} {96}},\ \bibinfo {pages} {134408}
  (\bibinfo {year} {2017})}\BibitemShut {NoStop}%
\bibitem [{\citenamefont {Franke}\ \emph {et~al.}(2022)\citenamefont {Franke},
  \citenamefont {C\ifmmode \u{a}\else \u{a}\fi{}lug\ifmmode~\u{a}\else
  \u{a}\fi{}ru}, \citenamefont {Nunnenkamp},\ and\ \citenamefont
  {Knolle}}]{FrankeKnolle_2022}%
  \BibitemOpen
  \bibfield  {author} {\bibinfo {author} {\bibfnamefont {O.}~\bibnamefont
  {Franke}}, \bibinfo {author} {\bibfnamefont {D.}~\bibnamefont {C\ifmmode
  \u{a}\else \u{a}\fi{}lug\ifmmode~\u{a}\else \u{a}\fi{}ru}}, \bibinfo {author}
  {\bibfnamefont {A.}~\bibnamefont {Nunnenkamp}}, \ and\ \bibinfo {author}
  {\bibfnamefont {J.}~\bibnamefont {Knolle}},\ }\href {\doibase
  10.1103/PhysRevB.106.174428} {\bibfield  {journal} {\bibinfo  {journal}
  {Phys. Rev. B}\ }\textbf {\bibinfo {volume} {106}},\ \bibinfo {pages}
  {174428} (\bibinfo {year} {2022})}\BibitemShut {NoStop}%
\bibitem [{Sun()}]{Sunny}%
  \BibitemOpen
  \href@noop {} {}\bibinfo {howpublished}
  {\url{https://github.com/SunnySuite/Sunny.jl}}\BibitemShut {NoStop}%
\bibitem [{Cod()}]{CodeExamples}%
  \BibitemOpen
  \href@noop {} {}\bibinfo {howpublished}
  {\url{https://github.com/SunnySuite/SunnyContributed}}\BibitemShut {NoStop}%
\bibitem [{\citenamefont {Swendsen}\ and\ \citenamefont
  {Wang}(1986)}]{PT_swedensen}%
  \BibitemOpen
  \bibfield  {author} {\bibinfo {author} {\bibfnamefont {R.}~\bibnamefont
  {Swendsen}}\ and\ \bibinfo {author} {\bibfnamefont {J.-S.}\ \bibnamefont
  {Wang}},\ }\href {\doibase 10.1103/PhysRevLett.57.2607} {\bibfield  {journal}
  {\bibinfo  {journal} {Physical review letters}\ }\textbf {\bibinfo {volume}
  {57}},\ \bibinfo {pages} {2607} (\bibinfo {year} {1986})}\BibitemShut
  {NoStop}%
\bibitem [{\citenamefont {Maksimov}\ \emph {et~al.}(2019)\citenamefont
  {Maksimov}, \citenamefont {Zhu}, \citenamefont {White},\ and\ \citenamefont
  {Chernyshev}}]{Maksimov_2019}%
  \BibitemOpen
  \bibfield  {author} {\bibinfo {author} {\bibfnamefont {P.}~\bibnamefont
  {Maksimov}}, \bibinfo {author} {\bibfnamefont {Z.}~\bibnamefont {Zhu}},
  \bibinfo {author} {\bibfnamefont {S.~R.}\ \bibnamefont {White}}, \ and\
  \bibinfo {author} {\bibfnamefont {A.}~\bibnamefont {Chernyshev}},\ }\href
  {\doibase 10.1103/PhysRevX.9.021017} {\bibfield  {journal} {\bibinfo
  {journal} {Physical Review X}\ }\textbf {\bibinfo {volume} {9}},\ \bibinfo
  {pages} {021017} (\bibinfo {year} {2019})}\BibitemShut {NoStop}%
\bibitem [{\citenamefont {Granroth}\ \emph {et~al.}(2010)\citenamefont
  {Granroth}, \citenamefont {Kolesnikov}, \citenamefont {Sherline},
  \citenamefont {Clancy}, \citenamefont {Ross}, \citenamefont {Ruff},
  \citenamefont {Gaulin},\ and\ \citenamefont {Nagler}}]{granroth2010}%
  \BibitemOpen
  \bibfield  {author} {\bibinfo {author} {\bibfnamefont {G.~E.}\ \bibnamefont
  {Granroth}}, \bibinfo {author} {\bibfnamefont {A.~I.}\ \bibnamefont
  {Kolesnikov}}, \bibinfo {author} {\bibfnamefont {T.~E.}\ \bibnamefont
  {Sherline}}, \bibinfo {author} {\bibfnamefont {J.~P.}\ \bibnamefont
  {Clancy}}, \bibinfo {author} {\bibfnamefont {K.~A.}\ \bibnamefont {Ross}},
  \bibinfo {author} {\bibfnamefont {J.~P.~C.}\ \bibnamefont {Ruff}}, \bibinfo
  {author} {\bibfnamefont {B.~D.}\ \bibnamefont {Gaulin}}, \ and\ \bibinfo
  {author} {\bibfnamefont {S.~E.}\ \bibnamefont {Nagler}},\ }\href {\doibase
  10.1088/1742-6596/251/1/012058} {\bibfield  {journal} {\bibinfo  {journal}
  {Journal of Physics: Conference Series}\ }\textbf {\bibinfo {volume} {251}},\
  \bibinfo {pages} {012058} (\bibinfo {year} {2010})}\BibitemShut {NoStop}%
\bibitem [{\citenamefont {Stone}\ \emph {et~al.}(2014)\citenamefont {Stone},
  \citenamefont {Niedziela}, \citenamefont {Abernathy}, \citenamefont
  {{DeBeer-Schmitt}}, \citenamefont {Ehlers}, \citenamefont {Garlea},
  \citenamefont {Granroth}, \citenamefont {{Graves-Brook}}, \citenamefont
  {Kolesnikov}, \citenamefont {Podlesnyak},\ and\ \citenamefont
  {Winn}}]{stone2014}%
  \BibitemOpen
  \bibfield  {author} {\bibinfo {author} {\bibfnamefont {M.~B.}\ \bibnamefont
  {Stone}}, \bibinfo {author} {\bibfnamefont {J.~L.}\ \bibnamefont
  {Niedziela}}, \bibinfo {author} {\bibfnamefont {D.~L.}\ \bibnamefont
  {Abernathy}}, \bibinfo {author} {\bibfnamefont {L.}~\bibnamefont
  {{DeBeer-Schmitt}}}, \bibinfo {author} {\bibfnamefont {G.}~\bibnamefont
  {Ehlers}}, \bibinfo {author} {\bibfnamefont {O.}~\bibnamefont {Garlea}},
  \bibinfo {author} {\bibfnamefont {G.~E.}\ \bibnamefont {Granroth}}, \bibinfo
  {author} {\bibfnamefont {M.}~\bibnamefont {{Graves-Brook}}}, \bibinfo
  {author} {\bibfnamefont {A.~I.}\ \bibnamefont {Kolesnikov}}, \bibinfo
  {author} {\bibfnamefont {A.}~\bibnamefont {Podlesnyak}}, \ and\ \bibinfo
  {author} {\bibfnamefont {B.}~\bibnamefont {Winn}},\ }\href {\doibase
  10.1063/1.4870050} {\bibfield  {journal} {\bibinfo  {journal} {Review of
  Scientific Instruments}\ }\textbf {\bibinfo {volume} {85}},\ \bibinfo {pages}
  {045113} (\bibinfo {year} {2014})}\BibitemShut {NoStop}%
\bibitem [{\citenamefont {Arnold}\ \emph {et~al.}(2014)\citenamefont {Arnold},
  \citenamefont {Bilheux}, \citenamefont {Borreguero}, \citenamefont {Buts},
  \citenamefont {Campbell}, \citenamefont {Chapon}, \citenamefont {Doucet},
  \citenamefont {Draper}, \citenamefont {Ferraz~Leal}, \citenamefont {Gigg},
  \citenamefont {Lynch}, \citenamefont {Markvardsen}, \citenamefont
  {Mikkelson}, \citenamefont {Mikkelson}, \citenamefont {Miller}, \citenamefont
  {Palmen}, \citenamefont {Parker}, \citenamefont {Passos}, \citenamefont
  {Perring}, \citenamefont {Peterson}, \citenamefont {Ren}, \citenamefont
  {Reuter}, \citenamefont {Savici}, \citenamefont {Taylor}, \citenamefont
  {Taylor}, \citenamefont {Tolchenov}, \citenamefont {Zhou},\ and\
  \citenamefont {Zikovsky}}]{arnold2014}%
  \BibitemOpen
  \bibfield  {author} {\bibinfo {author} {\bibfnamefont {O.}~\bibnamefont
  {Arnold}}, \bibinfo {author} {\bibfnamefont {J.}~\bibnamefont {Bilheux}},
  \bibinfo {author} {\bibfnamefont {J.}~\bibnamefont {Borreguero}}, \bibinfo
  {author} {\bibfnamefont {A.}~\bibnamefont {Buts}}, \bibinfo {author}
  {\bibfnamefont {S.}~\bibnamefont {Campbell}}, \bibinfo {author}
  {\bibfnamefont {L.}~\bibnamefont {Chapon}}, \bibinfo {author} {\bibfnamefont
  {M.}~\bibnamefont {Doucet}}, \bibinfo {author} {\bibfnamefont
  {N.}~\bibnamefont {Draper}}, \bibinfo {author} {\bibfnamefont
  {R.}~\bibnamefont {Ferraz~Leal}}, \bibinfo {author} {\bibfnamefont
  {M.}~\bibnamefont {Gigg}}, \bibinfo {author} {\bibfnamefont {V.}~\bibnamefont
  {Lynch}}, \bibinfo {author} {\bibfnamefont {A.}~\bibnamefont {Markvardsen}},
  \bibinfo {author} {\bibfnamefont {D.}~\bibnamefont {Mikkelson}}, \bibinfo
  {author} {\bibfnamefont {R.}~\bibnamefont {Mikkelson}}, \bibinfo {author}
  {\bibfnamefont {R.}~\bibnamefont {Miller}}, \bibinfo {author} {\bibfnamefont
  {K.}~\bibnamefont {Palmen}}, \bibinfo {author} {\bibfnamefont
  {P.}~\bibnamefont {Parker}}, \bibinfo {author} {\bibfnamefont
  {G.}~\bibnamefont {Passos}}, \bibinfo {author} {\bibfnamefont
  {T.}~\bibnamefont {Perring}}, \bibinfo {author} {\bibfnamefont
  {P.}~\bibnamefont {Peterson}}, \bibinfo {author} {\bibfnamefont
  {S.}~\bibnamefont {Ren}}, \bibinfo {author} {\bibfnamefont {M.}~\bibnamefont
  {Reuter}}, \bibinfo {author} {\bibfnamefont {A.}~\bibnamefont {Savici}},
  \bibinfo {author} {\bibfnamefont {J.}~\bibnamefont {Taylor}}, \bibinfo
  {author} {\bibfnamefont {R.}~\bibnamefont {Taylor}}, \bibinfo {author}
  {\bibfnamefont {R.}~\bibnamefont {Tolchenov}}, \bibinfo {author}
  {\bibfnamefont {W.}~\bibnamefont {Zhou}}, \ and\ \bibinfo {author}
  {\bibfnamefont {J.}~\bibnamefont {Zikovsky}},\ }\href {\doibase
  10.1016/j.nima.2014.07.029} {\bibfield  {journal} {\bibinfo  {journal}
  {Nuclear Instruments and Methods in Physics Research Section A: Accelerators,
  Spectrometers, Detectors and Associated Equipment}\ }\textbf {\bibinfo
  {volume} {764}},\ \bibinfo {pages} {156} (\bibinfo {year}
  {2014})}\BibitemShut {NoStop}%
\bibitem [{\citenamefont {Ewings}\ \emph {et~al.}(2016)\citenamefont {Ewings},
  \citenamefont {Buts}, \citenamefont {Le}, \citenamefont {Van~Duijn},
  \citenamefont {Bustinduy},\ and\ \citenamefont {Perring}}]{ewings2016}%
  \BibitemOpen
  \bibfield  {author} {\bibinfo {author} {\bibfnamefont {R.}~\bibnamefont
  {Ewings}}, \bibinfo {author} {\bibfnamefont {A.}~\bibnamefont {Buts}},
  \bibinfo {author} {\bibfnamefont {M.}~\bibnamefont {Le}}, \bibinfo {author}
  {\bibfnamefont {J.}~\bibnamefont {Van~Duijn}}, \bibinfo {author}
  {\bibfnamefont {I.}~\bibnamefont {Bustinduy}}, \ and\ \bibinfo {author}
  {\bibfnamefont {T.}~\bibnamefont {Perring}},\ }\href {\doibase
  10.1016/j.nima.2016.07.036} {\bibfield  {journal} {\bibinfo  {journal}
  {Nuclear Instruments and Methods in Physics Research Section A: Accelerators,
  Spectrometers, Detectors and Associated Equipment}\ }\textbf {\bibinfo
  {volume} {834}},\ \bibinfo {pages} {132} (\bibinfo {year}
  {2016})}\BibitemShut {NoStop}%
\bibitem [{\citenamefont {Cao}\ \emph {et~al.}(2018)\citenamefont {Cao},
  \citenamefont {Chakoumakos}, \citenamefont {Andrews}, \citenamefont {Wu},
  \citenamefont {Riedel}, \citenamefont {Hodges}, \citenamefont {Zhou},
  \citenamefont {Gregory}, \citenamefont {Haberl}, \citenamefont {Molaison},\
  and\ \citenamefont {Lynn}}]{cao2018}%
  \BibitemOpen
  \bibfield  {author} {\bibinfo {author} {\bibfnamefont {H.}~\bibnamefont
  {Cao}}, \bibinfo {author} {\bibfnamefont {B.}~\bibnamefont {Chakoumakos}},
  \bibinfo {author} {\bibfnamefont {K.}~\bibnamefont {Andrews}}, \bibinfo
  {author} {\bibfnamefont {Y.}~\bibnamefont {Wu}}, \bibinfo {author}
  {\bibfnamefont {R.}~\bibnamefont {Riedel}}, \bibinfo {author} {\bibfnamefont
  {J.}~\bibnamefont {Hodges}}, \bibinfo {author} {\bibfnamefont
  {W.}~\bibnamefont {Zhou}}, \bibinfo {author} {\bibfnamefont {R.}~\bibnamefont
  {Gregory}}, \bibinfo {author} {\bibfnamefont {B.}~\bibnamefont {Haberl}},
  \bibinfo {author} {\bibfnamefont {J.}~\bibnamefont {Molaison}}, \ and\
  \bibinfo {author} {\bibfnamefont {G.}~\bibnamefont {Lynn}},\ }\href {\doibase
  10.3390/cryst9010005} {\bibfield  {journal} {\bibinfo  {journal} {Crystals}\
  }\textbf {\bibinfo {volume} {9}},\ \bibinfo {pages} {5} (\bibinfo {year}
  {2018})}\BibitemShut {NoStop}%
\end{thebibliography}%
\end{document}